\documentclass[usenatbib,fleqn]{mnras}

\usepackage{mathptmx} 
\usepackage{amsmath} 
\usepackage{graphicx} 
\usepackage{hyperref} 
\usepackage{xspace} 
\usepackage{xcolor} 
\usepackage{upgreek} 
\usepackage[caption = false]{subfig} 
\usepackage{multirow} 

\renewcommand{\pi}{\uppi}

\newcommand{\Mpc}{\,h^{-1}\mathrm{Mpc}}

\newcommand{\iMpc}{\,h\mathrm{Mpc}^{-1}}
\newcommand{\eV}{\,\mathrm{eV}}
\newcommand{\Msun}{\,h^{-1}\mathrm{M_\odot}}
\newcommand{\Kelvin}{\,\mathrm{K}}

\newcommand{\Om}{\Omega_\mathrm{m}}

\newcommand{\Ob}{\Omega_\mathrm{b}}
\newcommand{\Ow}{\Omega_w}

\newcommand{\Oc}{\Omega_\mathrm{c}}

\newcommand{\om}{\omega_\mathrm{m}}
\newcommand{\ob}{\omega_\mathrm{b}}
 
\newcommand{\dc}{\delta_\mathrm{c}}
\newcommand{\Dv}{\Delta_\mathrm{v}}

\newcommand{\eg}{e.g.,\xspace}

\newcommand{\nbody}{$N$-body\xspace}
\newcommand{\LCDM}{$\Lambda$CDM\xspace}
\newcommand{\halofit}{\textsc{halofit}\xspace}

\newcommand{\hmcode}{\textsc{hmcode}\xspace}
\newcommand{\hmcodett}{\textsc{hmcode-2020}\xspace}
\newcommand{\cosmicemu}{\textsc{cosmic emu}\xspace}
\newcommand{\frankenemu}{\textsc{franken emu}\xspace}
\newcommand{\miratitan}{\textsc{mira titan}\xspace}
\newcommand{\Planck}{\textit{Planck}\xspace}
\newcommand{\WMAP}{\textit{WMAP}\xspace}
\newcommand{\kids}{KiDS\xspace}
\newcommand{\DES}{DES\xspace}
\newcommand{\HSC}{HSC\xspace}
\newcommand{\camb}{\textsc{camb}\xspace}
\newcommand{\class}{\textsc{class}\xspace}
\newcommand{\bahamas}{\textsc{bahamas}\xspace}
\newcommand{\owls}{\textsc{owls}\xspace}
\newcommand{\cosmoowls}{\textsc{cosmo-owls}\xspace}
\newcommand{\euclid}{\textsc{euclid}\xspace}
\newcommand{\bacco}{\textsc{bacco}\xspace}

\newcommand{\web}[1]{\href{#1}{#1}}

\newcommand{\hmcodelink}{\web{https://github.com/alexander-mead/HMcode}\xspace}
\newcommand{\VDliblink}{\web{http://powerlib.strw.leidenuniv.nl}\xspace}
\newcommand{\bahamaslink}{\web{http://www.astro.ljmu.ac.uk/~igm/BAHAMAS}\xspace}

\def\mnras{MNRAS}
\def\apj{ApJ}
\def\apjl{ApJ Let.}
\def\prd{Phys. Rev. D}
\def\aap{A\&A}
\def\jcap{J. Cosmol. Astropart. Phys.}
\def\apjs{ApJS}

\def\gtsim{\mathrel{\lower0.6ex\hbox{$\buildrel {\textstyle >}\over {\scriptstyle \sim}$}}}
\def\ltsim{\mathrel{\lower0.6ex\hbox{$\buildrel {\textstyle <}\over {\scriptstyle \sim}$}}}

\title[Improved non-linear power spectra]{\hmcodett: Improved modelling of non-linear cosmological power spectra with baryonic feedback}
\author[A. J. Mead et al.]{
A. J. Mead$^{1,2}$\thanks{alexander.j.mead@googlemail.com},
S. Brieden$^{2}$,
T. Tr\"oster$^{1}$,
 and
 C. Heymans$^{1,3}$
\\
1 Institute for Astronomy, University of Edinburgh, Royal Observatory, Blackford Hill, Edinburgh EH9 3HJ, UK
\\
2 Institut de Ci\`encies del Cosmos, Universitat de Barcelona, Mart\'i Franqu\`es 1, E08028 Barcelona, Spain
\\
3 Ruhr-University Bochum, Astronomical Institute, German Centre for Cosmological Lensing, Universit\"{a}tsstr. 150, 44801 Bochum, Germany
}

\pagerange{\pageref{firstpage}--\pageref{lastpage}} 
\pubyear{2020}

\begin{document}
\maketitle

\label{firstpage}

\begin{abstract}
We present an updated version of the \hmcode augmented halo model that can be used to make accurate predictions of the non-linear matter power spectrum over a wide range of cosmologies. Major improvements include modelling of BAO damping in the power spectrum and an updated treatment of massive neutrinos. We fit our model to simulated power spectra and show that we can match the results with an RMS error of $2.5$ per cent across a range of cosmologies, scales $k < 10\iMpc$, and redshifts $z<2$. The error rarely exceeds $5$ per cent and never exceeds $16$ per cent. The worst-case errors occur at $z\simeq2$, or for cosmologies with unusual dark-energy equations of state. This represents a significant improvement over previous versions of \hmcode, and over other popular fitting functions, particularly for massive-neutrino cosmologies with high neutrino mass. We also present a simple halo model that can be used to model the impact of baryonic feedback on the power spectrum. This six-parameter physical model includes gas expulsion by AGN feedback and encapsulates star formation. By comparing this model to data from hydrodynamical simulations we demonstrate that the power spectrum response to feedback is matched at the $<1$ per cent level for $z<1$ and $k<20\iMpc$. We also present a single-parameter variant of this model, parametrized in terms of feedback strength, which is only slightly less accurate. We make code available for our non-linear and baryon models at \hmcodelink and it is also available within \camb and soon within \class.
\end{abstract}

\begin{keywords}
cosmology: theory -- 
large-scale structure of Universe
\vspace{0.7cm} 
\end{keywords}

\section{Introduction}
\label{sec:introduction}

The non-linear power spectrum of matter fluctuations is a quantity of central importance in modern cosmology. This is primarily because weak-gravitational lensing \citep[\eg][]{Kilbinger2015} measures a projected version of the matter spectrum and therefore to extract accurate cosmological constraints from weak lensing observations requires accurate theoretical models of the power spectrum. When fluctuations are small, linear theory allows the power spectrum and its evolution to be calculated analytically -- an inflationary power-law spectrum of fluctuations is processed by linear electromagnetic and gravitational evolution \citep[\eg using \camb of][]{CAMB}. However, matter fluctuations inevitably become large as gravitational collapse generally amplifies inhomogeneity, and eventually linear perturbation theory \citep[\eg][]{Lewis2000, Blas2011}, and even non-linear perturbation theory \citep[\eg][]{Bernardeau2002}, fail to accurately track the details of the evolution of the power spectrum. Numerical simulations are currently the only known method for accurately calculating the deeply non-linear evolution of structure. However, simulations are expensive to run and it is therefore useful to have physically-motivated semi-analytical models of the non-linear regime of clustering as these can provide insight into an otherwise complex process. A second advantage of semi-analytical models is that they allow predictions to be made for cosmologies that have yet to be simulated.

The halo model \citep{Seljak2000, Peacock2000, Ma2000} is just such a semi-analytical model of clustering. All matter is considered to be contained within haloes and the clustering problem is then decomposed into two problems: first, the clustering that arises between haloes (two-halo) and second, the clustering that arises within individual haloes (one-halo). A halo-mass distribution function is required for the calculation, as well as a bias recipe for how haloes cluster with respect to matter, and a choice needs to be made for the halo profile. In its simplest form, the halo model makes a series of assumptions, all of which contribute to inaccuracies: haloes are assumed to be linearly biased tracers of an underlying linear matter density field with exactly spherical profiles the properties of which depend only on the halo mass. With reasonable choices, the model has been shown to be accurate for the matter--matter power spectrum at only the $\sim 30$ per-cent level compared to accurate spectra measured from simulations \citep[\eg][]{Tinker2005, Valageas2011, Mead2015b}. In addition, \cite{vanDaalen2015} showed that the underlying assumptions of the halo model cannot be correct in detail, as the non-linear power spectrum measured in simulations was shown not to be all accounted for by the matter within haloes. Despite this, the halo model is very widely used and has many uses beyond calculating the matter power spectrum, for example it can be used to calculate the power spectrum of galaxies \citep[\eg][]{Peacock2000}, the Sunyaev-Zel`dovich auto spectrum \citep[\eg][]{Komatsu2002}, the cross spectrum between galaxies and matter \citep[\eg][]{Cacciato2012, vandenBosch2013}, the cross spectrum between matter and electron pressure \citep[\eg][]{Ma2015, Mead2020a}, intrinsic alignments \citep[\eg][]{Fortuna2020} and higher-order statistics \citep[\eg][]{Cooray2002}. The halo model will be more or less accurate depending on the scenario to which it is applied. 

To counter the inaccuracy of the halo model for the matter spectrum \cite{Smith2003} developed the popular \halofit fitting function. A functional form was chosen, based on the halo model, with similar two- and one-halo terms, but called `quasi-linear' and `halo' terms. Parameters were fitted to simulated data over a wide range of cosmologies, specifically over the parameters $h$, $\sigma_8$, $n_\mathrm{s}$, $\Omega_\mathrm{m}$ and $\Omega_\Lambda$ (curvature included). The model was also fitted to numerical data from pure power-law simulations (in contrast to standard simulations, where the inflationary power law is modified by physical processes in the later universe to create a linear power spectrum with rich features) for theoretical consistency. \cite{Takahashi2012} updated the fitting function to account for inaccuracies that arose because of the limited resolution of simulations available when the original model was fitted, and also to provide a more accurate model for dark energy cosmologies with a fixed equation of state, governed by parameter $w$. Around the same time, \cite{Bird2012} provided a tweak to make the original \cite{Smith2003} model work for massive-neutrino cosmologies. These latter two approaches have never been officially combined, however, the popular \camb software contains an unpublished version of the \cite{Bird2012} neutrino model that works in tandem with the \cite{Takahashi2012} update. Recently, \cite{Smith2019} have provided an updated \halofit that combines perturbation theory at large scales with a small-scale prediction from a smoothing-spline-fit that characterises the differences between the \cite{Takahashi2012} fitting function and high-resolution simulations.

\begin{figure*}
\begin{center}
\hspace*{-0.35cm}\includegraphics[height=18cm,angle=270]{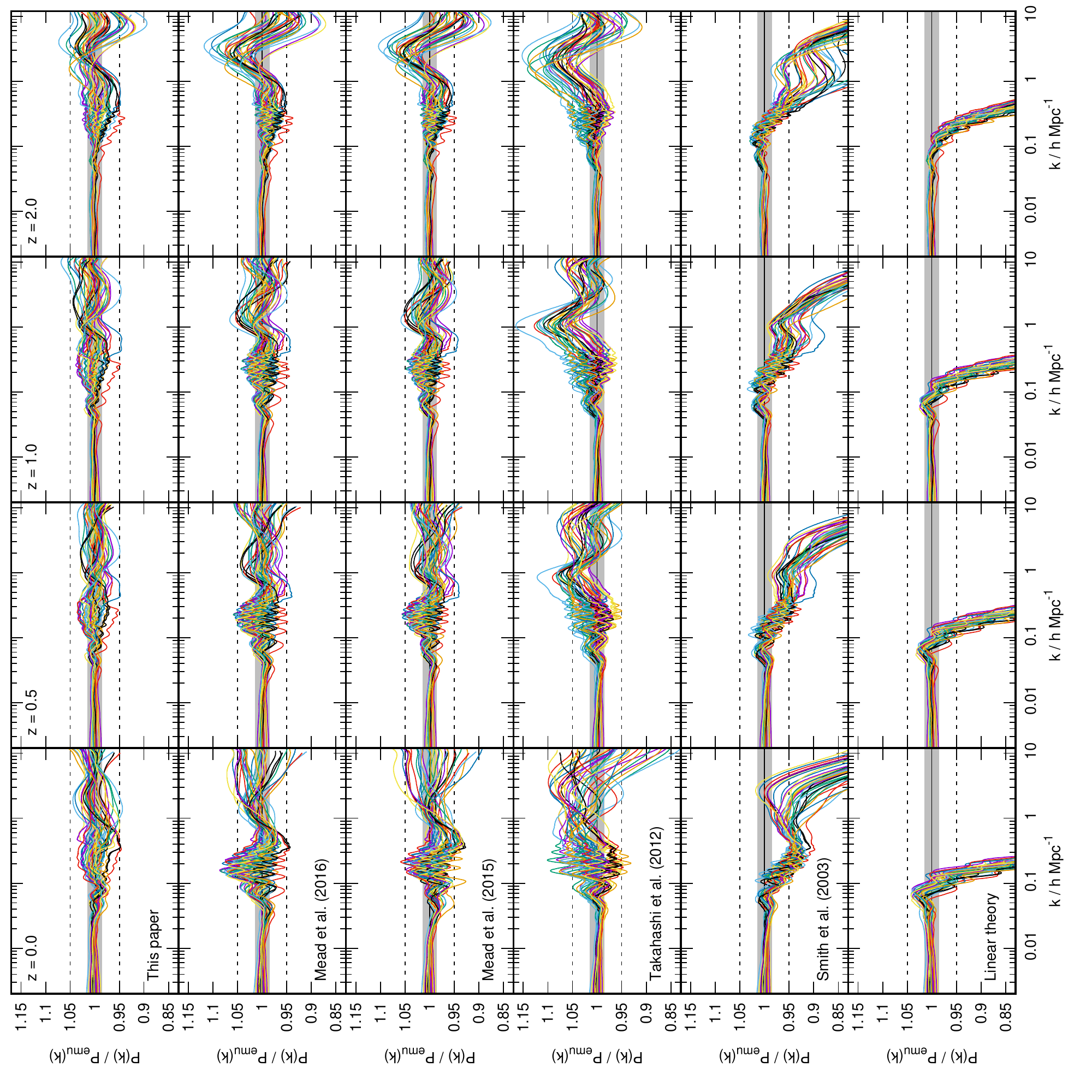}
\end{center}
\caption{Comparison of different matter power spectrum prediction schemes to the node cosmologies of the \frankenemu version of the cosmic emulator \protect{\citep{Heitmann2014}}. These cosmologies span a range of values of $\om$, $\ob$, $h$, $n_s$, $\sigma_8$ and $w$. Each coloured line shows a different cosmology and the columns show $z=0$ (left) $0.5$, $1$ and $2$ (right). The different rows show different prediction: \hmcodett presented in this paper (top), followed by \protect\cite{Mead2016}, and \protect\cite{Mead2015b}, \halofit from \protect\cite{Takahashi2012} followed by \protect\cite{Smith2003} and linear theory (bottom). Dashed horizontal lines show $5$ per cent errors, and accuracy (and publication date) increases from the bottom to the top. All the \hmcode models were fitted to data from these emulator nodes. The grey band, most obvious in the bottom two rows, indicates the $2\sigma$ region of the noise floor in the emulator predictions which we calculate from the scatter in the linear predictions are large scales. No prediction scheme can do better than this for these data.}
\label{fig:FrankenEmu}
\end{figure*}

\begin{figure*}
\begin{center}
\hspace*{-0.15cm}\includegraphics[height=18cm,angle=270]{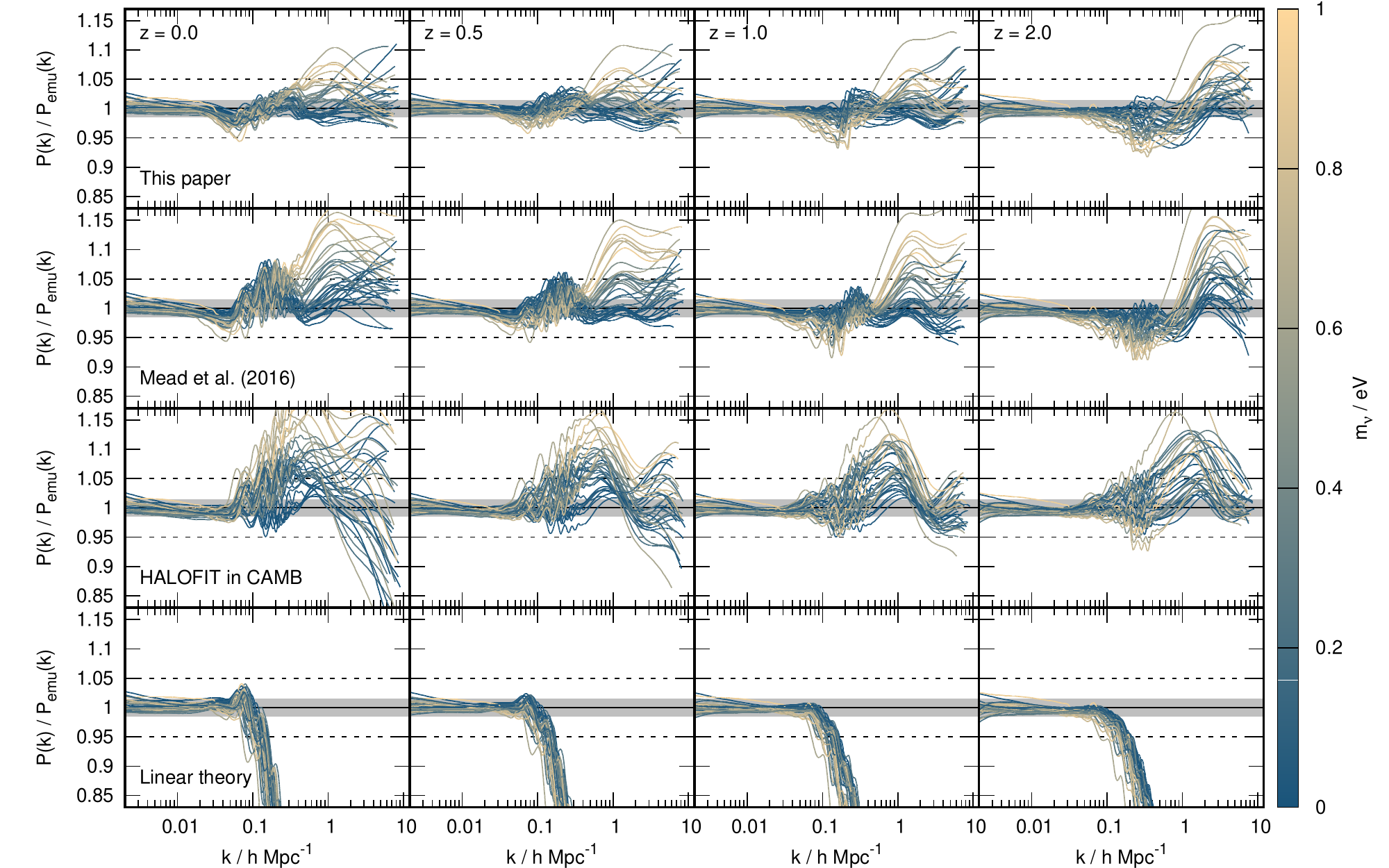}
\end{center}
\caption{As Fig.~\ref{fig:FrankenEmu} but for the nodes of the \miratitan version of the cosmic emulator. Compared to \frankenemu the emulator here additionally covers the extra parameters: time-varying dark energy equation of state, $w_a$; neutrino density, $\omega_\nu$. None of the models presented here were fitted to cosmologies from the nodes of this emulator, so this provides a fair comparison. Power spectra here are coloured by the value of neutrino mass so that the correlation of error with neutrino mass is clearly visible. Note that the minimal $m_\nu \simeq 0.06\eV$ corresponds almost to the bluest colour here and that the upper range probed by the emulator is quite high $\sim 1\eV$ compared to current constraints on the neutrino mass: $m_\nu < 0.12\eV$ \protect{\citep{Planck2018VI}}. Once again, the grey band indicates the $2\sigma$ region of the noise floor of the emulator node predictions.}
\label{fig:MiraTitan}
\end{figure*}

In \cite{Mead2015b} the original \hmcode\footnote{\cite{HMcode}: \hmcodelink} was presented, in which the standard halo-model calculation was augmented with fitted parameters so as to provide a better match to simulated matter power spectra data. \hmcode differs from \halofit in that the halo model calculation is the starting point for \hmcode, rather than a fitting function based on the halo model. The physical basis for these additional \hmcode fitted parameters is debatable, but it was shown that the \hmcode approach is more accurate than the \halofit approach, and that this accuracy can be achieved with fewer fitted parameters. \hmcode can be thought of as pertaining to an effective population of haloes whose power spectrum provides a better match to data. The hybrid nature of the model is an advantage because the grounding of the model in physical reality allows extensions to the standard cosmological paradigm to be easily considered \citep[][]{Mead2016}. In this paper we develop \hmcode to improve the accuracy so as to provide a tool that is appropriate for use with near-future cosmological data sets.

A comparison of some of the models for the matter power spectrum discussed in this introduction is shown in Figs.~\ref{fig:FrankenEmu} and \ref{fig:MiraTitan} where we show linear theory, \halofit of \cite{Smith2003}, \cite{Takahashi2012}, and the version of \cite{Takahashi2012} with the \cite{Bird2012} massive neutrino model, which we call `\halofit in \camb'. We also show \hmcode of \cite{Mead2015b}, \cite{Mead2016} as well as a preview of the model described in this paper. Each different curve shows the model compared to an \nbody simulated power spectrum\footnote{Taken from the node cosmologies of \cite{Heitmann2014} and \cite{Lawrence2017}} for a different cosmology. The date of publication and accuracy of model increases from the bottom to the top of the figure. These results will be discussed in detail in Section~\ref{sec:nonlinear_model}.

\begin{table}
\begin{center}
\caption{Ranges of cosmological parameters of the \frankenemu and \miratitan emulators that we use in this work. $\Omega_\mathrm{m}$, $\Omega_\mathrm{b}$ and $\Omega_\nu$ are the matter, baryon and neutrinos cosmological densities respectively; $\omega_i=\Omega_i h^2$, $n_\mathrm{s}$ is the spectral index of the inflationary spectrum, $\sigma_8$ is the RMS amplitude of linear fluctuations at $z=0$ in spheres of radius $8\Mpc$, $h$ is the dimensionless Hubble parameter, $w(a)=w_0+(1-a)w_a$ is the dark-energy equation of state. All cosmologies are flat. The node cosmologies of these emulators span the parameter ranges below using a Latin Hypercube design. Only \miratitan spans $w_a$ and $\omega_\nu$.}
\begin{tabular}{c c c c c}
\hline
\multirow{2}{*}{Parameter} &  \multicolumn{2}{c|}{\frankenemu} &  \multicolumn{2}{c|}{\miratitan} \\
\cline{2-5}
& min & max & min & max \\
\hline
$\omega_\mathrm{m}$ & $0.120$ & $0.155$ & $0.120$ & $0.155$ \\
$\omega_\mathrm{b}$ & $0.0215$ & $0.0235$ & $0.0215$ & $0.0235$ \\
$n_\mathrm{s}$ & $0.85$ & $1.05$ & $0.85$ & $1.05$ \\
$\sigma_8$ & $0.616$ & $0.9$ & $0.7$ & $0.9$ \\
$h$ & $0.55$ & $0.85$ & $0.55$ & $0.85$ \\
$w_0$ & $-1.3$ & $-0.7$ & $-1.3$ & $-0.7$ \\
$w_a$ & \multicolumn{2}{c}{-} & $-1.73$ & $1.28$ \\
$\omega_\nu$ & \multicolumn{2}{c}{-} & $0$ & $0.01$ \\
\hline
\end{tabular}
\label{tab:emulator_ranges}
\end{center}
\end{table}

The technique of `emulation' has been used to generate alternative theoretical models for the non-linear matter power spectrum by interpolating between the outputs of measurements from simulations. The original \cosmicemu was developed by \cite{Heitmann2010, Heitmann2009} and \cite{Lawrence2010} and report per-cent level accuracy for $k<1\iMpc$ over the parameter space of $\omega_\mathrm{m}$, $\omega_\mathrm{b}$, $n_\mathrm{s}$, $\sigma_8$ and $w$. \cosmicemu was extended by \cite{Heitmann2014} who used a combination of perturbation theory and extra simulations to extend the parameter space by including $h$ and to extend the range of scales covered to $k<10\iMpc$, at the expense of some accuracy, which was quoted at $4$ per cent after the Frankenstein-esque modifications: this approach was called \frankenemu. More recently \cite{Heitmann2016} and \cite{Lawrence2017} have created a new emulator, \miratitan, based on a new set of simulations and an improved emulation technique. They added the parameters $\omega_\nu$ and $w_a$ to the \frankenemu parameter space and quote an accuracy of $4$ per cent for $k<7\iMpc$, but plan to improve this in the future as more simulations are added to the hypercube. The ranges of \frankenemu and \miratitan are given in Table~\ref{tab:emulator_ranges}. As discussed by \cite{Harnois-Deraps2019}, the current range of the emulators is too tight to encompass the parameter space that is constrainable by current weak-lensing surveys such as \kids \citep[\eg][]{Hildebrandt2020, Asgari2020, Heymans2020}, \DES \citep[\eg][]{Abbott2018}, or \HSC \citep[\eg][]{Hikage2019, Hamana2020}, although they do contain the \cite{Planck2018VI} space. Other emulators of note include: an artificial-neural network, trained by \cite{Agarwal2012} and \cite{Agarwal2014}, which includes gas dynamics and massive neutrinos; the \textsc{euclid emulator} of \cite{Knabenhans2019, Knabenhans2020}; and that of \cite{Angulo2020}, which is based on a rescaling technique. \cite{Giblin2019} have proposed an emulator that would interpolate over linear power spectrum shape and evolution parameters, rather than cosmological parameters, as these have been demonstrated to be the primary drivers of structure formation. 

A confounding issue in the field of modelling the non-linear matter power spectrum is that of `baryonic feedback' \citep[\eg][]{Semboloni2011, Chisari2019b}: The models and emulators discussed previously were fitted or trained on data from `gravity-only' simulations, in which the effect of gas physics is ignored. In reality, non-gravitational processes such as gas heating as a byproduct of black-hole accretion, can inject large amounts of energy in galaxies and from there in to the surrounding halo. Hydrodynamical simulations \citep[\eg][]{Schaye2010, LeBrun2014, McCarthy2017} are needed, in conjunction with sub-grid recipes, to model these baryonic feedback processes. The physics of these processes are less well understood and more difficult to model compared to standard gravitational collapse. However, it is generally thought that gas expulsion from haloes lowers the amplitude of the power spectrum on scales that are relevant to weak-gravitational lensing observations. The details of this have been investigated via simulations \citep[\eg][]{Jing2006, vanDaalen2011, vanDaalen2014} and via the halo model \cite[\eg][]{White2004b, Zhan2004, Jing2006, Fedeli2014a, Fedeli2014b, vanDaalen2014, Debackere2020, Mead2020a}. In this paper we also provide a novel, physically-motivated model for how baryons affect the non-linear matter power.

This paper is organised as follows: In Section~\ref{sec:halo_model} we provide a short overview of the standard halo-model calculation for the matter power spectrum. In Section~\ref{sec:ingredients} we detail some tests that lead us to an optimal set of halo-model ingredients for our new version of \hmcode. In Section~\ref{sec:hmcode} we present the updated functional forms of \hmcodett and in Section~\ref{sec:nonlinear_model} we present the updated parameters for the matter power spectrum. In Section~\ref{sec:feedback_model} we provide a model for how baryonic feedback affects the power. Finally, we summarise in Section~\ref{sec:summary}. In Appendix~\ref{app:fitting_functions} we provide a fitting formula for spherical-collapse parameters that is valid for dark-energy and massive neutrino cosmologies. In Appendix~\ref{app:halofit_neutrinos} we discuss the error that different versions of \halofit make depending on the neutrino mass. In Appendix~\ref{app:early_dark_energy} we discuss the \hmcode error for cosmologies that have significant dark energy at early times. Appendix~\ref{app:emulators} shows comparisons between the non-linear power from \hmcodett and that from some popular emulators.

\section{Halo model}
\label{sec:halo_model}

In this Section we present a short review of the standard halo-model calculation \citep{Seljak2000, Peacock2000} for the matter power spectrum (sometimes we will refer to this as the matter--matter power spectrum, to emphasise that it is the spectrum of the matter field with itself). This calculation has been shown to be appropriate in cosmologies with massive neutrinos by \cite{Massara2014} for calculating the non-linear power spectrum of cold (non-neutrino) matter. The non-linear neutrino power spectrum was shown to be small in comparison, so we make the assumption that the non-neutrino matter accounts for all of the deeply non-linear power. We will interchange between $\Delta^2_{ij}(k,z)$ and $P_{ij}(k,z)$ power spectrum definitions:
\begin{equation}
\Delta^2_{ij}(k,z)=4\pi\left(\frac{k}{2\pi}\right)^3P_{ij}(k,z)\ .
\label{eq:power_definitions}
\end{equation}
The indices $i$ and $j$ pertain to the field: When considering massive neutrinos it is necessary to distinguish between the total matter field, denoted by `$\mathrm{m}$' and the cold (CDM and baryons) matter field, denoted by `$\mathrm{c}$', which differs from the total matter only in that the neutrinos have been subtracted\footnote{In principle $i$ and $j$ could be different. For example, $ij=\mathrm{mc}$ would indicate the cross spectrum between the total matter and the cold matter. However, in this paper we only use autospectra of total matter and cold matter, so $ij=\mathrm{mm}$ or $ij=\mathrm{cc}$, but we retain the two index notation for theoretical completeness.}. In cosmologies with no massive neutrinos the cold matter field is equal to the total matter field. As in \cite{Massara2014}, we define the overdensity of matter as $1+\delta_\mathrm{m}=\rho_\mathrm{m}/\bar\rho$ and the overdensity of cold matter as $1+\delta_\mathrm{c}=\rho_\mathrm{c}/\bar\rho_\mathrm{c}$. Defined in this way, the linear cold--cold power spectrum is equal to the linear matter--matter power spectrum at large scales, while it is \emph{greater} at smaller scales. For the spectra that we are interested in, $\Delta^2_{ij}(k,z)$ is dimensionless, while $P_{ij}(k,z)$ has units of volume. 

The non-linear matter power spectrum can be written as the sum of a two- and a one-halo term, respectively given by
\begin{equation}
P^{2\mathrm{H}}_\mathrm{mm}(k,z)=P^{\mathrm{lin}}_\mathrm{mm}(k,z)
\left[\int_0^\infty b(M,z)W(M,k,z)n(M,z)\;\mathrm{d}M\right]^2\ ,
\label{eq:two_halo}
\end{equation}
\begin{equation}
P^{1\mathrm{H}}_\mathrm{mm}(k,z)=\int_0^\infty W^2(M,k,z)n(M,z)\;\mathrm{d}M\ ,
\label{eq:one_halo}
\end{equation}
where $P^{\mathrm{lin}}_\mathrm{mm}(k,z)$ is the linear matter--matter power spectrum, $M$ is the halo mass, $b(M,z)$ is the linear halo bias, and $n(M,z)$ is the halo mass function: the distribution function for the number density of haloes in a mass range (sometimes denoted $\mathrm{d}n/\mathrm{d}M$ in the literature). Equations~(\ref{eq:two_halo}) and (\ref{eq:one_halo}) contain the (spherical) Fourier transforms of the halo matter density profiles:
\begin{equation}
W(M,k,z)=\frac{1}{\bar\rho}\int_0^{r_\mathrm{v}}4\pi r^2\frac{\sin(kr)}{kr}\rho(M,r,z)\;\mathrm{d}r\ ,
\label{eq:window_function}
\end{equation}
where $\rho(M,r,z)$ is the radial matter density profile in a host halo of mass $M$ and $\bar\rho$ is the mean matter density. The halo mass is related to the virial radius, $r_\mathrm{v}$ via
\begin{equation}
M = \frac{4}{3}\pi r_\mathrm{v}^3\Delta_\mathrm{v}(z)\bar\rho\ ,
\label{eq:virial_condition}
\end{equation}
where $\Delta_\mathrm{v}(z)$ is the virial halo overdensity. We lower the amplitude of $W(M, k, z)$ in the one-halo term by the factor $1-f_\nu$, where $f_\nu=\Omega_\nu/\Omega_\mathrm{m}$ is the neutrino mass fraction, to account for the fact that we assume that hot neutrinos cannot cluster in haloes and therefore do not contribute power to the one-halo term. Therefore $W(M,k\to0,z)=(1-f_\nu)M/\bar\rho$ and has units of volume. $M$ should be thought of as the mass the halo would have in an equivalent universe with no massive neutrinos (and no baryonic feedback, see Section~\ref{sec:feedback_model}).

The dimensionless mass function, $F(\nu,z)$, normalised such that the integral over all $\nu$ gives unity, is related to $n(M,z)$ via
\begin{equation}
F(\nu,z)\;\mathrm{d}\nu=\frac{M}{\bar\rho}n(M,z)\;\mathrm{d}M\ ,
\label{eq:mass_function}
\end{equation}
and is written in terms of the peak-height variable 
\begin{equation}
\nu=\dc(z)/\sigma_\mathrm{cc}(M,z)\ ,
\label{eq:peak_height}
\end{equation}
where $\dc(z)$ is the critical linear density threshold for halo collapse, commonly taken to be $1.686$. $\sigma_\mathrm{cc}(M,z)$ is the variance in the linear, cold matter field when filtered on a Lagrangian scale, $R$, corresponding to halo mass $M=4\pi R^3\bar\rho/3$. More generally we can consider the variance for any field pair $i$ and $j$:
\begin{equation}
\sigma^2_{ij}(R,z)=\int_0^{\infty}\Delta_{\mathrm{lin},ij}^2(k,z)\, T^2(kR)\;\mathrm{d}\ln{k}\ ,
\label{eq:variance_density}
\end{equation}
with $T(x)$ being the Fourier transform of a real-space top-hat filter
\begin{equation}
T(x)=\frac{3}{x^3}(\sin{x}-x\cos{x})\ .
\label{eq:top_hat}
\end{equation}
 \cite{Massara2014} showed that using the cold--cold matter variance in equation~(\ref{eq:peak_height}) produced better halo-model predictions for massive-neutrino cosmologies, while \cite{Castorina2014} showed that the cold definition produced a more universal halo bias and halo mass; we follow this here.

We find it useful to define the variance in the linear displacement field
\begin{equation}
\sigma_{\mathrm{v},ij}^2(R,z)=\frac{1}{3}\int_0^{\infty}\frac{\Delta_{\mathrm{lin},ij}^2(k,z)}{k^2}\, T^2(kR)\;\mathrm{d}\ln{k}\ ,
\label{eq:variance_displacement}
\end{equation}
where $\sigma_{\mathrm{v},ij}$ has units of distance and is finite in the $R\to0$ limit, unlike the variance in the overdensity in equation~(\ref{eq:variance_density}). To refer to this limit we use the notation $\sigma_{\mathrm{v},ij}(z)$, without an $R$ argument. The factor of $1/3$ converts from 3D to 1D displacement variance. We also find it useful to define the effective index of the power spectrum when smoothed on scale $R$, which is defined via
\begin{equation}
3+n^\mathrm{eff}_{ij}(R,z)=-\frac{\mathrm{d}\ln\sigma_{ij}^2(R,z)}{\mathrm{d}\ln R}\ .
\label{eq:neff}
\end{equation}
$n^\mathrm{eff}_{ij}(R,z)$ measures the slope of the linear power spectrum, and varies between $\sim n_\mathrm{s}$ for large $R$, to $\sim n_\mathrm{s}-4$ for small $R$. Here $n_\mathrm{s}$ is the spectral index of the linear power spectrum, and $n_\mathrm{s}\sim 1$ for reasonable cosmologies. Later we use $n^\mathrm{eff}_{ij}(R,z)$ evaluated at the non-linear scale, $R_\mathrm{nl}$ where $\sigma(R_\mathrm{nl},z)=\delta_\mathrm{c}(z)$, we denote this value by dropping the $R$ argument: $n^\mathrm{eff}_{ij}(z)$.

\section{Ingredient choices}
\label{sec:ingredients}

In Section~\ref{sec:hmcode} we add non-standard terms to the vanilla halo-model calculation and in Section~\ref{sec:nonlinear_model} we fit parameters in these in order to accurately match simulated data. However, before we do this we should determine a best-possible set of ingredients for our standard halo model, on which we base our modifications.  We must choose ingredients for: the  halo profiles (equation~$\ref{eq:window_function}$), the halo definition (equation~$\ref{eq:virial_condition}$) and the halo mass function (equation~$\ref{eq:mass_function}$). For each, there exist a myriad of possible choices in the literature, and often no clear metric for deciding on a best choice. In this Section, we show how we decided on a `best' choice of halo-model ingredients for \hmcode. This analysis is for the matter--matter power spectrum only, and it is possible that other ingredient choices are better for different power spectra \citep[\eg][]{Mead2020a}.

\begin{figure*}
\begin{center}
\hspace*{-0.6cm}\includegraphics[height=18cm,angle=270]{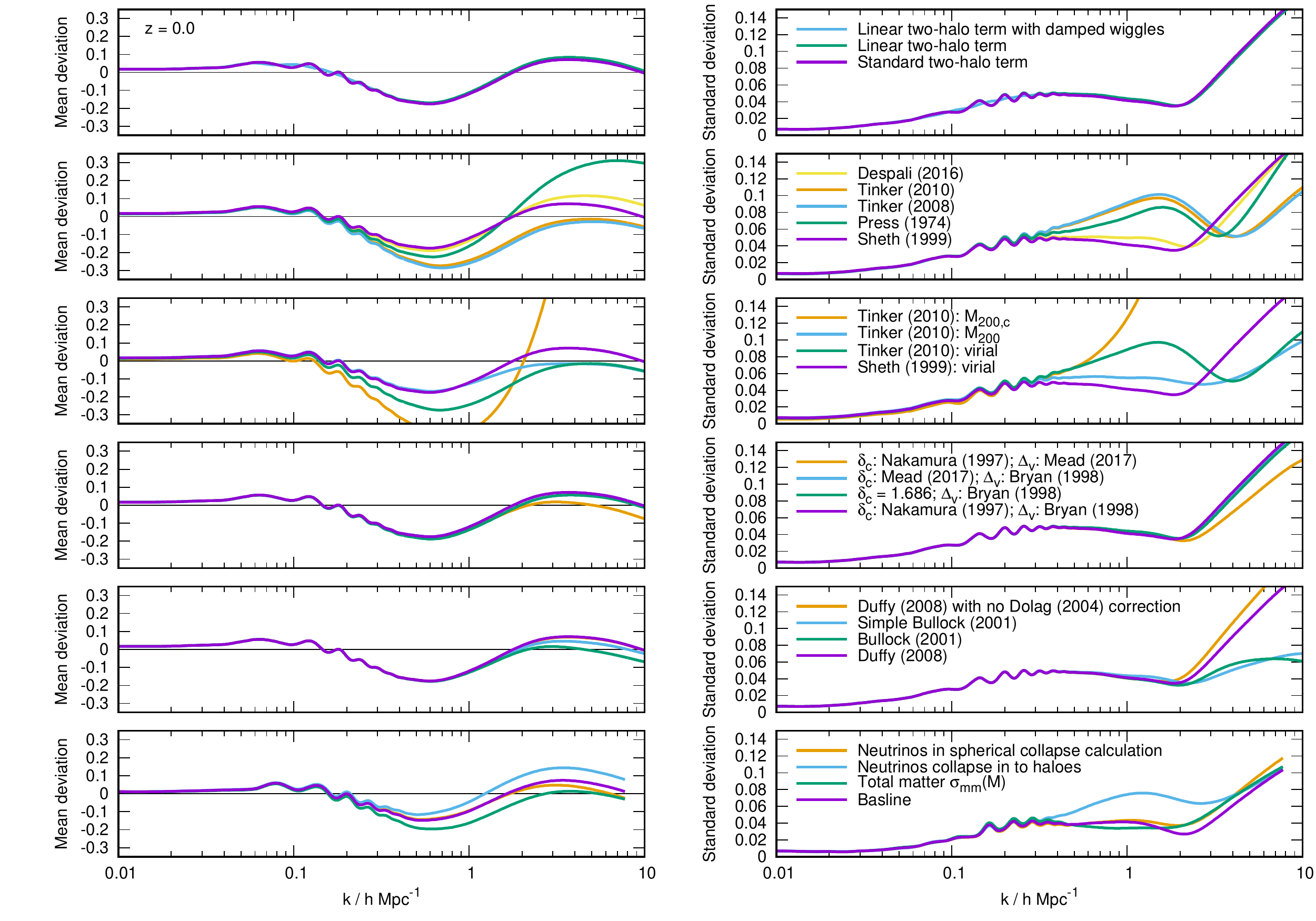}
\end{center}
\caption{Halo-model predictions for the matter--matter power for different choices of model ingredients. The top $5$ rows show comparison to the $37$ node cosmologies of \frankenemu. In the left-hand column we show the mean prediction across the different node cosmologies, while in the right-hand column we show the standard deviation of the prediction between the different cosmologies.  In each panel the purple curve shows the baseline model, from which all others deviate. The coloured curves in the  rows show different choices for the two-halo term (top), halo mass function, halo definition, spherical-collapse calculation and concentration--mass relation. The bottom row compares to the $26$ massive-neutrino node cosmologies of \miratitan, because massive neutrinos are not included in the cosmologies of \frankenemu. In this row ingredients that effect only the massive-neutrino implementation in the halo model are varied. In this Figure we show results at $z = 0$, but the results are broadly similar for other redshifts, although some of the trends are less extreme.}
\label{fig:emulator_mean_variance_all}
\end{figure*}

In Fig.~\ref{fig:emulator_mean_variance_all} we show how halo-model predictions at $z=0$ compare to the $37$ node cosmologies of the \frankenemu (or the $26$ massive-neutrino node cosmologies of \miratitan) for different choices of ingredients. The emulator nodes are discussed more thoroughly in Section~\ref{sec:nonlinear_model}. We show both the mean deviation
\begin{equation}
M(k) = \frac{1}{N} \sum_{i=1}^{N} \frac{P^{i}_\mathrm{mod}(k)-P^{i}_\mathrm{emu}(k)}{P^{i}_\mathrm{emu}(k)} \ ,
\end{equation}
and the standard deviation
\begin{equation}
S(k) = \sqrt{\frac{1}{N} \sum_{i=1}^{N} \left(\frac{P^{i}_\mathrm{mod}(k)-P^{i}_\mathrm{emu}(k)}{P^{i}_\mathrm{emu}(k)} - M(k)\right)^2}\ ,
\end{equation}
of the predictions, where the sums run over the $N$ cosmologies. A good ingredient should keep the mean deviation close to zero and minimize the standard deviation. For example, it would be unacceptable to choose an ingredient that worked very well for one specific cosmology but failed badly for other cosmologies. Similarly, if the mean prediction is good but the standard deviation between the predictions for different cosmologies is high we should be sceptical about the ingredient. In this Section, and throughout this paper, we always calculate our linear power spectrum using \camb \citep{CAMB}. We make deviations in choices of ingredients from a baseline halo model that uses: a virial definition of halo mass (equation~\ref{eq:virial_condition}), with the virial overdensity calculated using the fitting formula of \cite{Bryan1998}; the mass function and halo bias\footnote{We always calculate halo bias by applying the peak-background split formalism \citep{Mo1996,Sheth2001}.} from \cite{Sheth1999}\footnote{This mass function was fitted to simulated haloes defined using a virial halo definition.} with the linear collapse threshold calculated using the fitting formula of \cite{Nakamura1997}; we take \citeauthor{Navarro1997} (NFW; \citeyear{Navarro1997}) halo profiles with a concentration, $c=r_\mathrm{v}/r_\mathrm{s}$ appropriate for virial haloes taken from \cite{Duffy2008}. When we change halo definition we also update the concentration--mass relation to be consistent, again using \cite{Duffy2008}. Before discussing the merits of specific ingredient choices, the general trends in the mean and variance in Fig.~\ref{fig:emulator_mean_variance_all} show us a number of interesting things:
\begin{itemize}

\item
At the largest physical scales there is a non-zero variance of $\sim0.5$ per cent. This is the limiting intrinsic accuracy of the emulator, even at the node cosmologies. Assuming that this is not a function of scale gives us a limiting RMS error with $\sigma\simeq0.007$, beyond which we cannot push any model based on these data.

\item
There is a systematic positive bias in the mean at the largest scales, $k\simeq0.02\iMpc$, of magnitude $\sim 2$ per cent that is not improved by any choice of ingredient. This is caused by the unphysical contribution of the one-halo term at large scales. This problem diminishes at higher $z$. We address this issue in Section~\ref{sec:one_halo_damping}.

\item
There is a small ($\sim 3$ per cent on top of the previous $2$ per cent) over prediction of mean power at $k\simeq0.07\iMpc$, which is a consequence of linear theory not containing the initial `pre-virialization' dip in power, which is the first non-linear (one-loop) effect determined by perturbation theory \citep{Suto1991}. We address this issue in Section~\ref{sec:two_halo_damping}.

\item
There is a general and significant under-prediction of mean power ($\sim 20$ per cent) which peaks at $k\simeq0.7\iMpc$. This is the notorious transition region between the two- and the one-halo terms. No choice of ingredients dramatically ameliorates the lack of power in this region, which strongly suggests that a defect in the simplistic halo model is responsible for this error. On scales $k\sim0.3\iMpc$, slightly larger than the peak of the transition region, no choice of ingredient has an impact on the variance, which suggests that the missing part of the modelling has a significant cosmology dependence. We address these issues with a number of tweaks that are described in the next Section.

\item
On smaller scales, $k>1\iMpc$, the situation is improved, both in mean and variance, which suggests that a description of the power in the Universe originating from virialized haloes is more valid at these scales. It is for these scales that the mean and variance can be significantly improved by making a careful choice of ingredients.
\end{itemize}

We now inspect the different rows of Fig.~\ref{fig:emulator_mean_variance_all} in detail, and use this information to decide on an optimal set of ingredients upon which we base our augmented halo model.

\subsection{Two-halo term}
\label{sec:two_halo_term}
The top row of Fig.~\ref{fig:emulator_mean_variance_all} shows the effect on the power spectrum of different choices for the two-halo term: either simple linear theory, de-wiggled linear theory (discussed in Section~\ref{sec:BAO_damping}), or the full two-halo term (equation~\ref{eq:two_halo}). We see that the difference between using linear theory or the standard two-halo term is tiny for the matter--matter power spectrum. This is because these only differ on scales where the one-halo term dominates the total power spectrum. This implies that the choice of halo bias has an insignificant effect on the calculation for the matter--matter power. One the other hand, we see a slight improvement when we use de-wiggled linear theory, but only in the variance of the predictions at $k\sim 0.1\iMpc$. This improvement arises because some of the scatter in predictions that occurs when using linear theory primarily originates through the BAO not being damped, but this only contributes to the scatter since the mean of the wiggle is zero and the location of the wiggles are roughly uncorrelated between the different cosmologies. In previous versions of \hmcode we used linear theory for the two-halo term, but the results here lead us to choose de-wiggled linear theory for the \hmcodett two-halo term, and therefore to ignore the integral term that usually appears in equation~(\ref{eq:two_halo}), which is equivalent taking the $k\to0$ limit. Ignoring this integral has the advantage of approximately halving the computational time.

\subsection{Halo mass function}
In the second row of Fig.~\ref{fig:emulator_mean_variance_all} we show the differences that arise from choice of mass function, but all with a virial halo definition. We compare \cite{Sheth1999}, \cite{Press1974},  \cite{Tinker2008, Tinker2010}, and \cite{Despali2016}. We note that the \cite{Tinker2008} and \cite{Despali2016} mass functions do not satisfy the required normalisation condition for the mass function, but violating this condition does not present a problem for the one-halo term in practice because the integral (equation~\ref{eq:one_halo}) only depends on the properties of haloes in a tight mass range\footnote{For mass functions that do not satisfy the normalisation condition we force the large-scale normalisation of the two-halo term to be correct via the procedure described in the appendix of \cite{Mead2020a}.}. We see that the \cite{Despali2016} mass function provides the best mean for $k<1\iMpc$ but that it becomes considerably worse than the other choices at smaller scales. Unsurprisingly, \cite{Press1974} produces a poor mean and large variance while both the \cite{Tinker2008, Tinker2010} mass functions perform very similarly. Somewhat surprisingly, the \cite{Sheth1999} mass function slightly outperforms those of \cite{Tinker2008, Tinker2010} both in the mean prediction and in the variance at $k\sim1\iMpc$. This is despite the fact that \cite{Tinker2008, Tinker2010} has been demonstrated to be a better fit to the mass function data measured from simulations. The improvement in variance is presumably because \cite{Sheth1999} was calibrated to a wider range of cosmologies than \cite{Tinker2008, Tinker2010}. Based on this, we continue to use the \cite{Sheth1999} as the \hmcodett halo mass function, as in previous versions of \hmcode.

\subsection{Halo overdensity}
In the third row of Fig.~\ref{fig:emulator_mean_variance_all} we also show how the power spectrum predictions change when using the \cite{Tinker2010} mass function, but for different definitions of the halo boundary, either $200$ times the mean matter density, the virial definition, or $200$ times the critical density. This is possible for the \cite{Tinker2010} function because mass-function parameters are provided for a range of halo-mass definitions that can be interpolated between. We also change the halo concentration relation to that appropriate for the new definition using results from \cite{Duffy2008}. Using $M_{200,\mathrm{c}}$ is far less accurate, both in terms of mean and variance, compared to either $M_{200}$ or $M_\mathrm{vir}$. For \cite{Tinker2010}, using $M_{200}$ is slightly preferable to $M_\mathrm{vir}$, but the performance is still poorer for $k<3\iMpc$ than the model that adopts $M_\mathrm{vir}$ with the \cite{Sheth1999} mass function and it also only provides marginal benefits for the variance at smaller scales. This supports claims in \cite{Courtin2011}, \cite{Despali2016} and \cite{Mead2017} that a virial definition for haloes is preferable and is able to capture some of the complicated cosmology dependence of non-linear quantities.

In the fourth row we compare different choices for the calculation of the virial halo definition $\Dv$ and the critical-collapse threshold $\dc$. For $\Dv$ we compare using either the \cite{Bryan1998} or the \cite{Mead2017} fitting formulae, whereas for $\dc$ we compare fixing $1.686$, the \cite{Nakamura1997} fitting formula and the \cite{Mead2017} fitting formula. The \cite{Mead2017} fitting formulae for $\dc$ and $\Dv$ are similar in scope to their predecessors, but were fitted to spherical-collapse calculations for a wider range of cosmologies and are more accurate for dark-energy models -- they are detailed in Appendix~\ref{app:fitting_functions}. We see that choices of $\dc$ make very little difference to the overall mean prediction or to the standard deviation of predictions. However, there is an improvement in the mean and variance when we use \cite{Mead2017} for $\Dv$. For this reason we decide to use the virial definition and the \cite{Mead2017} fitting formulae for \hmcodett for both $\dc$ and $\Dv$.

\subsection{Halo concentration}
In the fifth row of Fig.~\ref{fig:emulator_mean_variance_all}, we show the impact of different concentration--mass relations for the NFW halo profile. We see that these only impact upon predictions at the comparatively small scale of $k>1\iMpc$, which makes sense since these changes only affect the innermost portions of the halo. Using either of the two concentration--mass relations presented in \cite{Bullock2001} provide similarly good mean predictions, however, using the more complicated model minimizes the variance at the smallest scales. Either model suppresses the variance dramatically compared to using concentration--mass relations from \cite{Duffy2008}, which makes sense because the \cite{Bullock2001} relations are expressed as in terms of cosmology-dependent quantities, such as $\sigma(M,z)$ or $M_*(z)$, defined via $\sigma(M_*,z)=\delta_\mathrm{c}(z)$, whereas the relations of \cite{Duffy2008} are expressed as simplistic power-laws as functions of halo mass. It is clear that the concentration--mass relation has cosmology dependence, and so a relation that includes this is to be preferred. We also see an improvement in the variance when we apply the \cite{Dolag2004} correction (present in our baseline model) for dark-energy cosmologies. We therefore choose to use the full \cite{Bullock2001} relation with a \cite{Dolag2004} correction in \hmcodett, exactly as in previous versions of \hmcode.

\subsection{Massive neutrino choices}
Finally, in the bottom row of Fig.~\ref{fig:emulator_mean_variance_all} we show the impact of different choices regarding the implementation of massive neutrinos in our halo model. The baseline case is that the variance that appears in the mass function (equation~\ref{eq:peak_height}) is calculated using the cold--cold spectrum, neutrino mass is removed from the one-halo term (discussed below equation~\ref{eq:virial_condition}), and that neutrino mass is counted as matter in spherical-collapse calculations of $\Delta_v(z)$ and $\delta_c(z)$ (equations~\ref{eq:virial_condition} and \ref{eq:peak_height}). We show this baseline case as well as what happens when each of this assumptions is modified in turn: calculating the variance using the matter--matter spectrum; ignoring the $(1-f_\nu)$ correction to halo mass, therefore assuming that neutrinos cluster along with cold matter; accounting for massive neutrinos in the calculation of $\delta_\mathrm{c}$ and $\Delta_\mathrm{v}$ via the \cite{Mead2017} fitting functions (see Appendix~\ref{app:fitting_functions}). We see that the mean deviation and variance are minimized when we take $\sigma_\mathrm{cc}(M)$ and keep the factors of $(1-f_\nu)$ both of these choices are in agreement with  with results presented in \cite{Massara2014}. We also see an improvement when we use the \cite{Mead2017} relations for $\dc$ and $\Dv$ that have a massive-neutrino dependence, thus we choose this for \hmcodett. 

\section{HMcode 2020}
\label{sec:hmcode}

Following \cite{Mead2015b, Mead2016}, in this Section we detail some non-standard features that we are forced to add to the halo model power spectrum calculation in order to return a good match to the simulation data across a wide-range of cosmologies. While these tweaks are not from first principles, we find that they are essential in order to get a good match to power spectra measured in simulations. These parameters make up for inherent deficiencies in the standard halo-model calculation, some of which have been addressed by other authors: beyond-linear perturbation theory \citep[][]{Smith2007, Valageas2011, Schmidt2016, Philcox2020}; non-linear halo bias \citep[][]{Smith2007, Mead2021}; halo exclusion \citep[][]{Takada2004, Smith2011a, Cacciato2012, vandenBosch2013}; profile compensation \citep[][]{Cooray2002, Chen2019}; aspherical haloes \citep{Smith2005}; scatter in halo properties at fixed mass \citep[\eg][]{Giocoli2010}; matter not contained in haloes \citep[][]{Smith2011b, Voivodic2020}; halo substructure \citep[][]{Sheth2003, Giocoli2010}. We do not include any of these advanced strategies in the present work because they can considerably increase the computation time of a numerical halo-model implementation; one of the prime advantages of \hmcode is that the calculation is fast. In Section~\ref{sec:nonlinear_model} we fit our model to the node cosmologies of the \frankenemu and \miratitan emulators.

\subsection{BAO damping}
\label{sec:BAO_damping}

The first addition we make is to explicitly add baryon-acoustic oscillation (BAO) damping to the halo model. It is well known that non-linear gravitational evolution smears out the initially clear BAO signal. In configuration space this manifests as a decrease in amplitude and a broadening of the BAO peak in the correlation function. In Fourier space this manifests as a scale-dependent decrease in amplitude of the `wiggle' part of the power spectrum \citep[\eg][]{Crocce2006a, Taruya2010, Senatore2015}. To calculate de-wiggled linear theory we split the linear power spectrum into the sum of a smooth broadband component, $P_\mathrm{smt}(k,z)$, and a BAO wiggle,\footnote{To create the smoothed power spectrum we first divide the linear spectrum by the `no-wiggle' approximation from \cite{Eisenstein1998} and we then smooth the resulting function of $\ln(k/h\,\mathrm{Mpc}^{-1})$ with a Gaussian with $\sigma=0.25$. The eventual de-wiggled power is quite insensitive to $\sigma$, and our results are minimally affected for values between $0.15$ and $0.35$.} $P_\mathrm{wig}(k,z)$:
\begin{equation}
P^\mathrm{lin}_\mathrm{mm}(k,z)=P_\mathrm{smt}(k,z)+ P_\mathrm{wig}(k,z)\ .
\end{equation}
 We then damp the wiggle using the standard result for perturbation theory \citep[\eg][]{Crocce2006a} before returning the wiggle back to the smooth power spectrum to create the de-wiggled linear spectrum,
\begin{equation}
P_\mathrm{dwl}(k,z)=P^\mathrm{lin}_\mathrm{mm}(k,z) - (1-\mathrm{e}^{-k^2\sigma_\mathrm{v,mm}^2(z)})P_\mathrm{wig}(k,z)\ .
\end{equation}
This procedure involves no additional free parameters. 

\subsection{Perturbative damping of the two-halo term}
\label{sec:two_halo_damping}

To form our two-halo term we also damp the de-wiggled power spectrum, so that
\begin{equation}
P^{2\mathrm{H}}_\mathrm{mm}(k,z) = P_\mathrm{dwl}(k,z)\left[1-f\frac{(k/k_\mathrm{d})^{n_\mathrm{d}}}{1+(k/k_\mathrm{d})^{n_\mathrm{d}}}\right]\ .
\label{eq:two_halo_damping}
\end{equation}
As long as $n_\mathrm{d}>0$, the multiplicative term in square brackets is unity for $k\ll k_\mathrm{d}$ and $(1-f)$ for $k\gg k_\mathrm{d}$. Perturbation theory tells us that the largest-scale non-linear effect is a slight (few per cent at $k\sim 0.03\iMpc$ at $z=0$) damping of power that arises because voids become larger, but grow less quickly, than linear theory would predict, which is what this term aims to encapsulate. This introduces the free parameters $f$, $n_\mathrm{d}$ and $k_\mathrm{d}$. Note that the form of the damping in equation~(\ref{eq:two_halo_damping}) is different to that in the previous versions of \hmcode. Using equation~(\ref{eq:two_halo_damping}) for our two-halo term means that we ignore the integral terms that appear in equation~(\ref{eq:two_halo}), effectively setting them to unity, as justified in Section \ref{sec:two_halo_term}.

\subsection{Large-scale damping of the one-halo term}
\label{sec:one_halo_damping}

As shown in \cite{Smith2003}, as a consequence of mass and momentum conservation, the one-halo term should grow as $P_{1\mathrm{H}}(k)\propto k^4$ at large scales, rather than being constant. This feature is absent in the standard halo-model calculation because of a failure to account for negative perturbations (only positive-density haloes are considered; \citealt{Cooray2002} and \citealt{Chen2019} discuss profile compensation as a way round this). We impose this restriction on the one-halo term via the ad-hoc modification
\begin{equation}
P^{1\mathrm{H}}_\mathrm{mm}(k,z)\to P^{1\mathrm{H}}_\mathrm{mm}(k,z)\times\frac{(k/k_*)^4}{1+(k/k_*)^4}\ ,
\label{eq:one_halo_damping}
\end{equation}
which suppresses the (otherwise constant) one-halo term at large scales but leaves it unchanged at small scales. This term introduces the single free parameter $k_*$. If a suppression is not imposed, the one-halo term can unphysically contribute power on large scales ($k\ltsim0.01\iMpc$ at $z=0$) where linear theory is known to be near perfect. In previous versions of \hmcode a different form for one-halo damping was used, with exponential damping, but we prefer the current form since it respects the physical $P_\mathrm{1H}\propto k^4$ large-scale limit.

\subsection{One-halo ingredients}

We adopt the mass function of \cite{Sheth1999}:
\begin{equation}
F(\nu)\mathrm{d}\nu=A\left[1+\frac{1}{(q\nu^2)^p}\right]\mathrm{e}^{-q\nu^2/2}\;\mathrm{d}\nu\ ,
\label{eq:ST_mass_function}
\end{equation}
where $\nu$ is the peak height, defined in equation~(\ref{eq:peak_height}), and the normalisation $A$ is constrained by the condition that $F(\nu)\mathrm{d}\nu$ must integrate to unity. Standard values are $p=0.3$, $q=0.707$, which gives $A\simeq 0.216$. Note that the mass function in equation~(\ref{eq:ST_mass_function}) has no explicit cosmology or redshift dependence, which is all encoded in the transformation $\nu=\delta_\mathrm{c}(z)/\sigma(M,z)$. We take $\delta_\mathrm{c}(z)$ to be given by the fitting formula in \cite{Mead2017}, which is detailed in Appendix~\ref{app:fitting_functions} and which accurately reproduces spherical-collapse calculations for a wide range of cosmologies. This reduces to the standard $\simeq 1.686$ when $\Om= 1$, but is more accurate than the fitting formula of \cite{Nakamura1997} for dark-energy cosmologies.

As in \cite{Mead2015b,Mead2016}, we take our halo profiles to be a modified form of those from \citeauthor*{Navarro1997} (NFW; \citeyear{Navarro1997})
\begin{equation}
\rho_\mathrm{c}(M,r)\propto\frac{1}{r/(\nu^\eta r_\mathrm{s})[1+r/(\nu^\eta r_\mathrm{s})]^2}\ ,
\label{eq:modified_NFW_profile}
\end{equation}
with $r_\mathrm{s}$ related to $r_\mathrm{v}$ by the concentration parameter: $c=r_\mathrm{v}/r_\mathrm{s}$. The profile is assumed to extend out to $\nu^\eta r_\mathrm{v}$, and to be zero beyond that. Here $\eta$ is a new free `halo bloating' parameter; $\eta =0$ would correspond to the standard NFW profile. As shown in \cite*{Copeland2020}, the factors of $\nu^\eta$ in equation~(\ref{eq:modified_NFW_profile}) correspond to the Fourier-space change $W(M,k,z)\to W(M,\nu^\eta k,z)$ applied to the halo window functions in equation~(\ref{eq:one_halo}). The normalisation of the NFW profile is calculated from equation~(\ref{eq:virial_condition}), ensuring that the halo mass is enclosed within the (modified) virial radius. We take $\Delta_\mathrm{v}(z)$ from the spherical-collapse fitting function provided by \cite{Mead2017}, which is similar to that given in \cite{Bryan1998}, but more accurate for dark energy cosmologies. The formula is detailed in Appendix~\ref{app:fitting_functions}. 

We take our halo mass--concentration relation to be that of \cite{Bullock2001}
\begin{equation}
c(M,z)=B\left[\frac{1+z_\mathrm{f}(M,z)}{1+z}\right]\ ,
\label{eq:concentration_mass}
\end{equation}
where the formation redshift, $z_\mathrm{f}$ is calculated from
\begin{equation}
\frac{g(z_\mathrm{f})}{g(z)}\sigma_\mathrm{cc}(\gamma M,z) = \delta_\mathrm{c}(z)\ .
\label{eq:formation_redshift}
\end{equation}
This is essentially answering the question of: `at what redshift did the halo attain a fraction $\gamma$ of its final mass at $z$?' $\gamma = 0.01$ is taken, exactly as in \cite{Bullock2001}, even though it is strange for the concentration to be determined when such a small fraction of the mass has accumulated. $g(z)$ is the linear growth function, which is discussed in Appendix~\ref{app:fitting_functions}. For cosmologies that have a scale-dependent linear growth, we calculate a linear growth function in the large-scale limit with neutrinos clustered along with CDM. When solving equation~(\ref{eq:formation_redshift}) it is possible to find $z_\mathrm{f}<z$, in this case we follow \cite{Bullock2001} and force $z_\mathrm{f}=z$ so that the halo concentration can never be below the value $B$, which we take as a free parameter.

We modify the concentration--mass relation so as to be appropriate for dark-energy cosmologies using the prescription of \cite{Dolag2004}, whereby we modify the concentration by the ratio of linear growth factors at the halo `collapse' redshift
 \begin{equation}
 c(M,z) \to c(M, z) \frac{g(z_\mathrm{c})}{g_\Lambda(z_\mathrm{c})}\frac{g_\Lambda(z)}{g(z)}\ .
 \label{eq:Dolag_correction}
 \end{equation}
$g_\Lambda(z)$ is the growth factor calculated in an equivalent \LCDM cosmology where we force the dark energy to be $\Lambda$ (fixing $w=-1$ and $w_a=0$), enforce flatness (fixing $\Omega_\Lambda = 1 - \Om$) and convert any neutrino mass into CDM mass. The final fraction on the right-hand side of equation~(\ref{eq:Dolag_correction}) was not present in previous \hmcode versions, or in \cite{Dolag2004}, but we consider it necessary for the correction to make sense in the high-$z$ limit when the dark-energy density is too small to have an appreciable effect on the formation of haloes. \cite{Dolag2004} showed that evaluating their correction at the halo-collapse redshift was almost equivalent to evaluating it at an infinite redshift, so when evaluating the formula numerically we take $z_\mathrm{c}=10$ to be suitably infinite. Note that in \cite{Mead2016} the correction in equation~(\ref{eq:Dolag_correction}) was strengthened via an additional $1.5$ exponent, but we do not find this to be necessary in our present work.

\subsection{Total power}

As in \cite{Mead2015b, Mead2016} we find it necessary to modify the transition between the two- and the one-halo terms when constructing the full halo-model power. We do this via
\begin{equation}
\Delta^2_\mathrm{HM}(k,z)=\left[[\Delta^2_\mathrm{2H}(k,z)]^\alpha+[\Delta^2_\mathrm{1H}(k,z)]^\alpha \right]^{1/\alpha}\ ,
\label{eq:full_power}
\end{equation}
where $\alpha = 1$ would correspond to a standard transition. $\alpha<1$ smooths the transition while $\alpha>1$ sharpens it. We treat the parameter $\alpha$ as free.

\section{Non-linear model}
\label{sec:nonlinear_model}

\subsection{Parameter fitting}

The free parameters in our model are constrained using the 37 node cosmologies of the \frankenemu emulator, which reports to be accurate at the $4$ per-cent level. However, this is the quoted accuracy for the emulation scheme, and evaluating at the node cosmologies is considerably more accurate given that this involves no interpolation. From Fig.~\ref{fig:FrankenEmu} we estimate the accuracy to be $\ltsim 1$ per cent at the nodes. The set of node cosmologies spans a Latin Hypercube, with parameter ranges given in Table~\ref{tab:emulator_ranges}. Working with the nodes has the additional advantage of allowing us to exploit the Hypercube design. The parameter space of \frankenemu encompasses $\om$, $\ob$, $n_\mathrm{s}$, $\sigma_8$, $w$ and $h$.

\begin{table*}
\begin{center}
\caption{\hmcodett parameters. We list the parameter, with a short explanation of the meaning of the parameter, the equation where the parameter is defined in the text and the default value. We then show the best-fitted value, which in most cases is a functional form, and then an example of this function evaluated (to 3 significant figures) at $z=0$ for a standard \LCDM ($\Om=0.3$, $\sigma_8=0.8$) cosmology. There are a total of 12 fitted parameters.}
\begin{tabular}{c c c c c c}
\hline
Parameter & Explanation & Equation & Default value & Fitted functional form or value & Example $z=0$ value (3 s.f.) \\
\hline
$k_\mathrm{d}$ & Two-halo term damping wavenumber & \ref{eq:two_halo_damping} & - & $0.05699\times\sigma_{8,\mathrm{cc}}^{-1.089}(z)\iMpc$ & $0.073\iMpc$ \\
$f$ & Two-halo term fractional damping & \ref{eq:two_halo_damping} & $0$ &  $0.2696\times\sigma_{8,\mathrm{cc}}^{0.9403}(z)$ & $0.219$ \\
$n_\mathrm{d}$ & Two-halo term damping power & \ref{eq:two_halo_damping} & - & $2.853$ & $2.85$ \\
$k_*$ & One-halo term damping wavenumber & \ref{eq:one_halo_damping} & $0$ & $0.05618\times\sigma_{8,\mathrm{cc}}^{-1.013}(z)\iMpc$ & $0.070\iMpc$ \\
$\eta$ & Halo bloating & \ref{eq:modified_NFW_profile} & 0 & $0.1281\times\sigma_{8,\mathrm{cc}}^{-0.3644}(z)$ & $0.139$ \\
$B$ & Minimum halo concentration & \ref{eq:concentration_mass} & $4$ & $5.196$ & $5.20$ \\
$\alpha$ & Transition smoothing & \ref{eq:full_power} & $1$ & $1.875\times(1.603)^{n^\mathrm{eff}_\mathrm{cc}}$ & $0.719$ \\
\hline
\end{tabular}
\label{tab:fitted_parameters}
\end{center}
\end{table*}

Our best-fit parameters are given in Table~\ref{tab:fitted_parameters}. To perform the fitting we used the \cite{Nelder1965} simplex algorithm with a large number of initial starting points. We constrain the parameters using all node cosmologies, with an equal logarithmic weight in $k$ between $0.01$ and $10\iMpc$ and using data from $z=0$, $0.5$, $1$ and $2$ with equal weight. Initially cosmologies and redshifts were fitted independently, taking constant values for parameters, to determine if the model worked well for the cosmologies on a case-by-case basis. We then fitted the cosmologies and redshifts simultaneously, and plot the residuals to determine reasonable functional forms to encompass cosmology and redshift dependence. We parametrize this via quantities that depend on the power spectrum shape and amplitude, rather than the classic cosmological parameters. This is sensible because the Universe does not really \emph{know} the value of cosmological parameters like $h$, $\Om$, $w$ etc., instead structure formation is driven by the spectral shape and amplitude, and evolution \citep[][]{Peacock1994, Smith2003}. Our functional forms thus depend on $\sigma_{8,\mathrm{cc}}(z)$ and $n^\mathrm{eff}_\mathrm{cc}(z)$ (equations~\ref{eq:variance_density} and \ref{eq:neff}). It is important to note that all the cosmology dependence of parameters in our model depend on the \emph{cold} spectrum, as this is the quantity that drives structure formation in massive-neutrino cosmologies. Recall that the cold power spectrum is equal to the matter power spectrum in the absence of massive neutrinos.

\subsection{Model comparison}

\begin{table*}
\begin{center}
\caption{Accuracy of different predictions schemes for the non-linear matter power spectrum compared to the node cosmologies of \frankenemu and \miratitan. The number of fitted parameters is given, along with the additional number of fitted parameters for massive-neutrino models. The first number in the accuracy columns show the RMS residual of the model with respect to the emulator averaged logarithmically over $k$ (between $0.01$ and $10\iMpc$ for \frankenemu and between $0.01$ and $7\iMpc$ for Mira Titan) and over $z=0$, $0.5$, $1$ and $2$. The second number in the accuracy column shown the maximum error over the same range over all node cosmologies.} 
\begin{tabular}{c c c c c c c}
\hline
\multirow{2}{*}{Prediction scheme} & \multirow{2}{*}{Reference} & \multirow{2}{*}{\# fitted parameters} &  \multicolumn{2}{c}{\frankenemu error [\%]} & \multicolumn{2}{c}{\miratitan error [\%]} \\
\cline{4-7}
& & & RMS & max & RMS & max \\
\hline
\hmcodett & This paper & 12 & 1.7 & 10.6 & 2.5 & 15.9 \\
\hmcode & \cite{Mead2016} & 12 (+2) & 2.3 & 12.9 & 4.2 & 22.9 \\
\hmcode & \cite{Mead2015b} & 12 & 2.2 & 12.5 & \multicolumn{2}{c}{-} \\ 
\halofit & \cite{Takahashi2012}, \cite{Bird2012} & 34 (+6) & 3.6 & 16.5 & 5.0 & 24.7 \\
\halofit & \cite{Smith2003} & 30 & 9.5 & 43.5 & \multicolumn{2}{c}{-} \\
\hline
\end{tabular}
\label{tab:accuracy}
\end{center}
\end{table*}

Our model is shown compared to the 37 node cosmologies of \frankenemu in Fig.~\ref{fig:FrankenEmu} and then compared to the 36 node cosmologies of \miratitan in Fig.~\ref{fig:MiraTitan}, alongside comparisons with other popular fitting functions. In the \frankenemu comparison we show linear theory, \halofit from \cite{Smith2003} and \cite{Takahashi2012} and \hmcode from \cite{Mead2015b} and \cite{Mead2016}. In the \miratitan figure we omit comparisons with \cite{Smith2003} and \cite{Mead2015b} since these models were not fitted to cosmologies that included massive neutrinos. In Fig.~\ref{fig:MiraTitan} we colour the cosmologies according to the neutrino mass in order to show the correlation with the error. The accuracy of all \hmcode and \halofit versions are summarised in Table~\ref{tab:accuracy}. In Appendix~\ref{app:emulators} we show additional comparisons between \hmcodett, some other emulators, and the recent version of \halofit from \cite{Smith2019}.

In the bottom row of Figs.~\ref{fig:FrankenEmu} and \ref{fig:MiraTitan} we see a comparison of linear theory (from \camb, which we also use in \hmcodett) against the emulators. We see that linear theory provides a good description of the data at large scales, but we see that it starts to underestimate the true power spectrum at smaller scales. How well linear theory matches the non-linear result is a function of both $z$ and cosmology, as would be expected, with linear theory providing a better match to smaller scales at higher $z$. It is perhaps surprising that we see a small, per-cent level, scatter even at the largest scales shown, when we know that linear theory should be a near-perfect \citep[][]{Lesgourgues2011a} description of the power spectrum. This scatter arises due to the emulation scheme not perfectly reproducing the results at the node cosmologies, and so this gives an indication of how accurately the emulators work in the best-case scenario. If we take the error to be scale independent, then the emulator predicts the true power for the node cosmologies with an accuracy better than $1$ per cent.

Fig.~\ref{fig:FrankenEmu} shows the performance of the two different versions of \halofit, the original of \cite{Smith2003} and the update of \cite{Takahashi2012}, compared to the \frankenemu node cosmologies. We see that the original \cite{Smith2003} model systematically under-predicts the power spectrum for $k > 0.1\iMpc$ to a level that reaches tens of per cent by $k=10\iMpc$. The \halofit of \cite{Takahashi2012} performs much better, with errors around the 10 per cent level for the range of scales and redshifts we show. The general under-prediction of power at small scales of the original \halofit was one of the issues addressed by \cite{Takahashi2012} by using simulations with higher resolution. In Fig.~\ref{fig:MiraTitan} we show the comparison with \miratitan, which contains cosmologies with time-varying dark energy equations of state, $w(a)=w_0+(1-a)w_a$, so we have to make a choice if we interpret the \cite{Takahashi2012} `$w$' as $w_0$ or $w(a)$. We chose the latter based on the fact that this produces better results. The same choice is made in the implementation of \cite{Takahashi2012} within \camb. We show results with an unpublished version of the \cite{Bird2012} neutrino model applied to the \cite{Takahashi2012} \halofit, as this is the default \halofit prediction given by \camb, this is discussed in Appendix~\ref{app:halofit_neutrinos}. In Fig.~\ref{fig:MiraTitan} we can see that the \halofit error degrades when massive neutrinos and time-varying dark energy are added. There is a noticeable correlation of the error with neutrino mass at $z>0$. In both figures there is a significant scatter around the BAO scale and a further scatter for $k\sim10\iMpc$ at $z=0$. At higher redshifts there is a systematic `kink' feature in the residuals at $k\sim1\iMpc$.We find that the RMS accuracy of the \cite{Takahashi2012} version of \halofit is $4.2$ per cent over both sets of emulator nodes, but that errors can be as much as $24.7$ per cent for some cosmologies.

Fig.~\ref{fig:FrankenEmu} also shows the performance of the \cite{Mead2015b} and \cite{Mead2016} versions of \hmcode compared to the node cosmologies of \frankenemu. Both \hmcode versions were fitted to the node cosmologies of \frankenemu, so these results present a best-case scenario for the models. We see that the performance of the two versions are almost identical, with an RMS error of $2.3$ per cent and the error rarely exceeding $5$ per cent. The performance degrades at $k\simeq10\iMpc$ at $z=2$ compared to the other scales and redshifts. Fig.~\ref{fig:MiraTitan} shows the performance of \hmcode to the \miratitan nodes. The \cite{Mead2015b} version is omitted from this figure as it was not designed to work with cosmologies that include massive neutrinos. We see that for low neutrino mass the performance is similar to the case of \frankenemu, which is reassuring given that this version of \hmcode was not fitted to \miratitan data. However, we see a significant error for higher neutrino masses which manifests as a systematic under-prediction of power for $k\simeq 0.1\iMpc$ followed by a significant and systematic over-prediction of power for $k\simeq 1\iMpc$. For \miratitan the RMS error of this version of \hmcode is $4.2$ per cent with errors as large as $22.9$ per cent for the most extreme neutrino masses. The neutrino model of \hmcode was only fitted to data for $m_\nu<0.6\eV$, so it is perhaps not surprising that it fails more significantly for higher neutrino masses. These results are in accordance with those presented in \cite{Lawrence2017}.

\hmcodett can be seen in the top rows of Fig.~\ref{fig:FrankenEmu} (\frankenemu) and Fig.~\ref{fig:MiraTitan} (\miratitan). The model was fitted \frankenemu, but not to \miratitan. The mean error between our new model from the data is $1.7$ per cent for \frankenemu and we can see that the error rarely exceeds $5$ per cent, with it being around $10$ per cent in the worst-case scenario ($z=2$, $k\simeq 10\iMpc$). This represents a significant improvement over all previous fitting functions for the non-linear matter--matter power spectrum. For \miratitan, which is not used in our fitting, the results are degraded, but we still have an RMS error of $2.5$ per cent and a worst-case error of $15.9$ per cent, both of which are significantly better than the other models. As for the other versions of \hmcode, the performance degrades slightly at $z=2$ compared to the other redshifts. This is perhaps surprising given that one may assume non-linear structure formation to be simpler at higher redshift given that structure is less developed. However, this may be because the assumptions in the halo model of virialized spherical haloes may be less valid at higher redshift, when structure is forming more rapidly due to the shape of standard linear power spectra\footnote{At early times, many scales cross the threshold for collapse simultaneously because $\Delta^2$ is roughly constant at the relevant scales. At later times $\Delta^2$ is steeper and structure formation is more orderly.}. We also see that there is some correlation of the error with neutrino mass, with the model systematically over-predicting power for $k\simeq1\iMpc$ at $z=0$ and systematically under-predicting power for $k\simeq0.2\iMpc$ at $z=2$ for the highest neutrino masses. We also note that for a few cosmologies the model fails more severely for $k>1\iMpc$. As shown in Appendix~\ref{app:early_dark_energy}, these specific cosmologies all have early-time dark energy equations of state that are close to zero, and therefore qualify as `early dark energy' cosmologies. These equations of state means that the dark-energy density decays very slowly with redshift compared to other models; for example $\Ow(z=100) \sim 0.1$. We suspect that cosmologies like these may have significantly different structure formation pathways, and that our prescription for calculating the halo concentration may not be correct for these exotic cosmologies.

\subsection{Comments on \hmcode parameters}

\begin{figure}
\begin{center}
\includegraphics[height=8.5cm,angle=270]{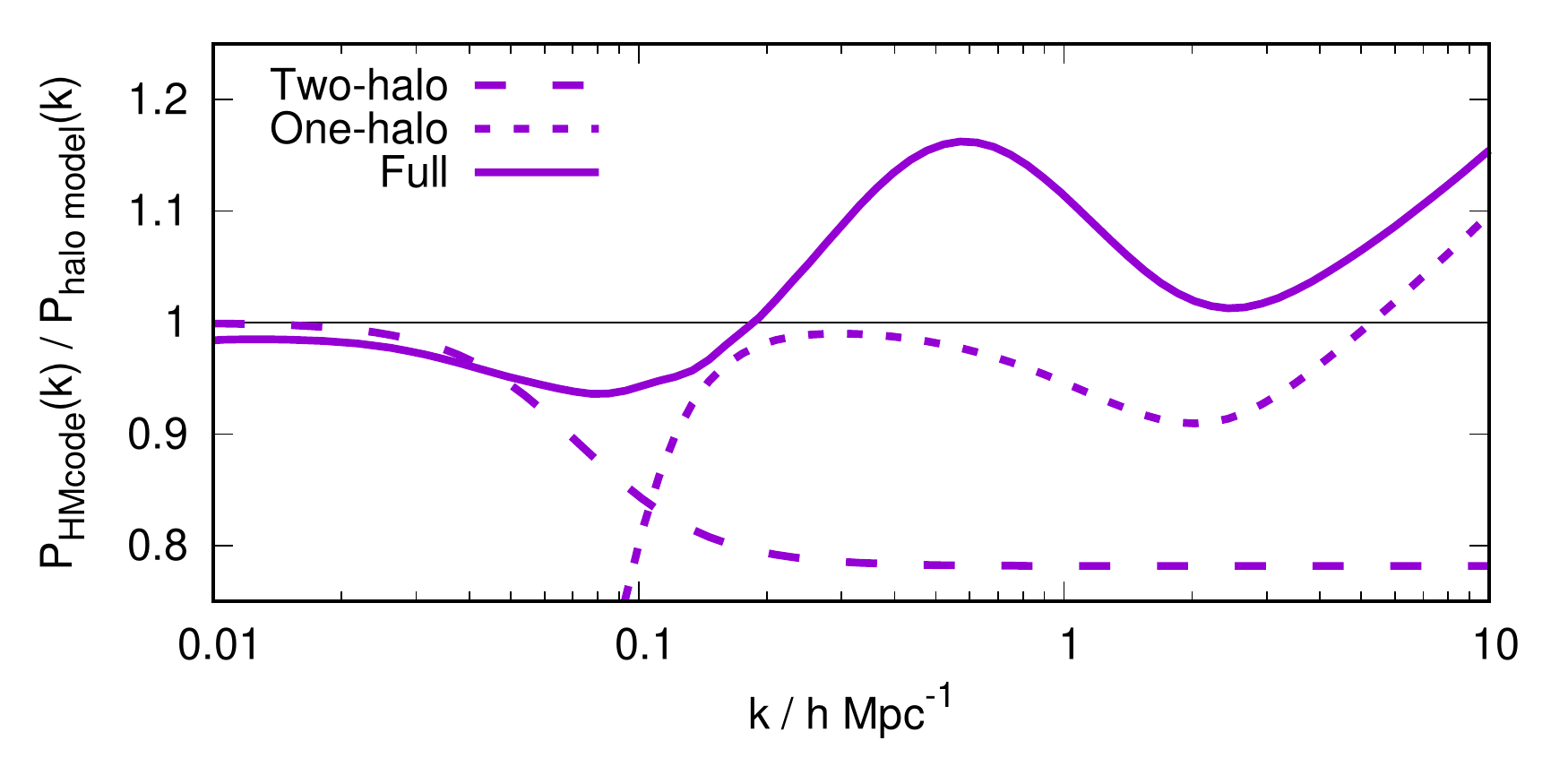}
\end{center}
\caption{Comparison of the ratio of power spectra for \hmcodett compared to that from the regular halo-model calculation with the same basic ingredients. The models are shown for a standard \LCDM cosmology at $z=0$. The two-halo term (long dashed), the one-halo term (short-dashed), and the total (solid) are shown for each model to illustrate the main differences between them.}
\label{fig:HMcode_halomod_comparison}
\end{figure}

We now comment on the values of some of the \hmcode fitted parameters shown in Table~\ref{tab:fitted_parameters}: 
\begin{itemize}

\item 
The value of $f$ at $z=0$ in a standard \LCDM cosmology is $0.22$ with $k_\mathrm{d}=0.073\iMpc$. Referring to equation~(\ref{eq:two_halo_damping}), this is perhaps surprising since we anticipated that this parameter would account for few per-cent (not $20$ per cent) effects at $\simeq 0.03\iMpc$, which would cover perturbative effects at these scales. However, this damping does have per-cent level effects on the power at $k=0.04\iMpc$ so it feasibly does account for this damping, but it has a much larger effect at smaller scales than we anticipated ($\simeq 10$ per cent at $k=0.1\iMpc$). From Fig.~\ref{fig:FrankenEmu} we can see that linear theory already under-predicts the power by around $10$ per cent at this scale so adding another $10$ percent deficit at this scale without any compensation would be catastrophic. The required compensation arises from the $\alpha$ parameter (equation~\ref{eq:full_power}) that is smoothing the transition between the two- and one-halo term, such that part of the one-halo term adds power back at these scales.


\item 
The wavenumbers for the two-halo damping and one-halo damping are very similar, both $\simeq 0.07\iMpc$. Given the very different roles that these damping terms play we can only assume this to be coincidental.

\item 
As in previous versions of \hmcode, we find it necessary to include the halo bloating parameter $\eta$, which deforms haloes in a mass-dependent way. The $z=0$ value for a standard \LCDM cosmology is $0.139$, the fact that this is positive makes $\nu^\eta$ in equation~(\ref{eq:modified_NFW_profile}) greater than unity for $\nu>1$ haloes and visa versa. This means that high-mass haloes are larger than they would otherwise be; $\simeq16$ per cent larger for a $\nu=3$ ($\sim10^{15}\Msun$ at $z=0$ for standard \LCDM) halo. A combination of this halo bloating, the concentration modification, and the transition region smoothing seem to be necessary to get a good match to the non-linear power spectrum by creating a one-halo term that is more curved than the standard. 

\item
The value of $\eta$ is closer to the default than it was in previous versions of \hmcode (where $\eta\simeq0.4$). This is probably because our default halo model (Section~\ref{sec:ingredients}), on which \hmcode is based, is a better description of the data prior to any parameter fitting.

\end{itemize}

Finally, in Fig.~\ref{fig:HMcode_halomod_comparison} we show a comparison between the power from the standard halo-model calculation described in Section~\ref{sec:ingredients}, and the fitted \hmcode as described in this Section. We see: the one-halo term at the largest scales falling off more steeply than standard (equation~\ref{eq:one_halo_damping}); the small-scale suppression in the two-halo term, for $k\sim0.1\iMpc$ (equation~\ref{eq:two_halo_damping}); the boost in power in the transition region ($k\sim0.5\iMpc$), we can infer that this is predominantly due to the $\alpha$ parameter (equation~\ref{eq:full_power}) because this boost is not present in either the two- or the one-halo terms; and a boost of power at $k\sim10\iMpc$ due to the modified concentration--mass relation.

\section{Baryonic feedback}
\label{sec:feedback_model}

In the previous Section we developed a model for the matter power spectrum assuming that gravity is the only important force for structure formation. This assumption is built in to our model given that we use ingredients for the halo model, such as the mass function and halo profiles, that are calibrated on simulations that only consider the gravitational interaction, and also because we have fitted our model to power spectra from such simulations. However, in reality we know that electromagnetic processes, particularly those associated with star formation and black-hole accretion, can have a significant impact on the distribution of matter. In this Section we develop a simple model to account for the effect of these baryonic feedback processes on the matter power spectrum. This is possible with the halo model since we have physical information such as the masses, distribution, and structural properties of haloes that we expect to be altered by feedback.

\subsection{Baryonic feedback model}

We parametrize an effective model for baryonic physics by including three physically-motivated changes to the standard halo model (\emph{not} \hmcodett) discussed in Section~\ref{sec:ingredients}: 

\begin{itemize}

\item
We allow feedback to deform haloes via a change in halo concentration \citep[][]{Rudd2008} via the parameter $B$ in equation~(\ref{eq:concentration_mass}). This is similar to the approach taken in \cite{Mead2015b} and \cite{Mead2020a}. Physically we expect that gas expulsion from haloes removes mass from the halo centre, thus lowering the effective concentration from the default $B=4$.

\item
We include a central delta-function term, of magnitude $f_*$, in the halo density profile to account for the presence of stars within haloes. As shown in \cite{Fedeli2014a}, \cite{Debackere2020} and \cite{Mead2020a}, a term like this is necessary to model the power spectra of stellar matter as seen in hydrodynamic simulations. Stars that have an appreciable effect on the matter power spectrum predominantly cluster in the centres of haloes and this creates a shot-noise term in the power spectrum, as well a cross term between this and the NFW profile, both of which contribute to additional small-scale power. The parameter $0<f_*<\Ob/\Om$ can be thought of as an effective halo stellar mass fraction. 

\item
We account for gas expulsion by lowering the gas content of haloes via:
\begin{equation}
f_\mathrm{g}(M)=\left(\frac{\Ob}{\Om}-f_*\right)\frac{(M/M_\mathrm{b})^\beta}{1+(M/M_\mathrm{b})^\beta}\ ,
\label{eq:gas_fraction}
\end{equation}
where $f_\mathrm{g}$ is the halo gas fraction, the pre-factor in parenthesis is the available gas reservoir, while $M_\mathrm{b}>0$ and $\beta>0$ are fitted parameters. Haloes of $M\gg M_\mathrm{b}$ are unaffected while those of $M< M_\mathrm{b}$ have lost more than half of their gas.

\end{itemize}

To implement these changes, we replace the NFW window function, $W(M,k)$, that would normally enter equation~(\ref{eq:one_halo}), with
\begin{equation}
\tilde{W}(M,k) = \left[\frac{\Oc}{\Om}+f_\mathrm{g}(M)\right]W(M,k)+f_*\frac{M}{\bar\rho}\ .
\label{eq:feedback_model}
\end{equation}
In the $M\ll M_\mathrm{b}$ limit halo masses are lowered by the fraction $\Omega_\mathrm{c}/\Omega_\mathrm{m}+f_*$, while in the opposite limit they are altered by $1-f_\nu$ as in the gravity-only case. The removal of gas mass implied by equation~(\ref{eq:gas_fraction}) lowers the overall amplitude of the one-halo term as well as changes its shape. 


In previous versions of \hmcode the feedback model was more basic. The parameters $B$ from equation~(\ref{eq:concentration_mass}) and $\eta$ from equation~(\ref{eq:modified_NFW_profile}) were fitted to data from the original \owls simulations \citep{Schaye2010, vanDaalen2011} to provide a model that approximately matched the suppression due to AGN feedback. Both this change in halo concentration and this `halo bloating' were found to be necessary to provide a good match. However, there was no term to account for star formation, so the model would only ever predict a suppression in power

\subsection{Baryonic feedback parameters}

The free parameters in our feedback model are constrained using data from the hydrodynamical library\footnote{\VDliblink} of \cite{vanDaalen2020}, which contains simulations from the \cosmoowls \citep{LeBrun2014} and \bahamas\footnote{\bahamaslink} \citep{McCarthy2017} suites. We choose to fit the power spectrum `response', as advocated in \cite{Mead2017}, \cite{Cataneo2019} and \cite{Mead2020a}\footnote{\hmcodett matches the \textsc{dmonly} spectra from the library at the $\simeq2.5$ per-cent level for $k<20\iMpc$, which is within the expected range based on Figs.~\ref{fig:FrankenEmu} and \ref{fig:MiraTitan}.}. For the simulations, the response is the matter--matter power spectrum measurement divided by that measured in an equivalent `gravity only' simulation. This approach has the advantage that we cancel out Gaussian variance at large scales in the simulations. For the halo model response we take the feedback model described in the previous subsection and divide it by the model described in Section~\ref{sec:ingredients}. We fit for the free parameters $B$ (equation~\ref{eq:concentration_mass}), $M_\mathrm{b}$ (equation~\ref{eq:gas_fraction}), and $f_*$ (equation~\ref{eq:feedback_model}), and we allow for redshift dependence of the form
\begin{equation}
X(z) = X_0\times 10^{z X_z}\ ,
\label{eq:feedback_model_z}
\end{equation}
where $X$ can be either $B$, $M_\mathrm{b}$, or $f_*$ and there are two free parameters, $X_0$ and $X_z$, for each $X$. This functional form was determined by initially fitting the model independently of $z$ and then examining the preferred trends in fitted parameters. We fix $\beta=2$ in equation~(\ref{eq:feedback_model}) since there was no clear preference for a specific value, and this worked well for all simulations considered. The eventual feedback model therefore has $6$ parameters. Redshifts were fitted simultaneously from $z=0$ to $1$ with a linear weight, and wavenumbers were fitted with a logarithmic weight from $k=0.03$ to $20\iMpc$. 

\begin{table*}
\begin{center}
\caption{Best-fitting parameter values for our $6$-parameter feedback model that match the data from \bahamas and \cosmoowls hydrodynamical simulations. The exact cosmological parameters of each simulation can be found in table 2 of \protect\cite{vanDaalen2020}. For the \WMAP9 simulations with neutrinos the value of $\sigma_8$ decreases as the neutrino mass increases. For the \Planck2015 simulations with neutrinos all cosmological parameters change as the neutrino mass changes such that the \Planck CMB data remain well fitted.} 
\begin{tabular}{c c c c c c c c c c c}
\hline
Simulation suite & Feedback & $\log_{10}(T_\mathrm{AGN}/\Kelvin)$ & Cosmology & $M_\nu/\eV$ & $B_0$ & $B_z$ & $f_{*,0}/10^{-2}$ & $f_{*,z}$ & $\log_{10}(M_{\mathrm{b},0}/\Msun)$ & $M_{\mathrm{b},z}$ \\
\hline
\bahamas & \textsc{agn} & $7.6$ & \WMAP9 & 0 & $3.55$ & $-0.060$ & $2.06$ & $0.40$  & $13.52$ & $-0.13$ \\
\bahamas & \textsc{agn} & $7.8$ & \WMAP9 & 0 & $3.41$ & $-0.066$ & $2.03$ & $0.41$  & $13.84$ & $-0.14$ \\
\bahamas & \textsc{agn} & $8.0$ & \WMAP9 & 0 & $3.37$ & $-0.075$ & $1.94$ & $0.41$ &  $14.24$ & $-0.05$ \\
\hline
\bahamas & \textsc{agn} & $7.8$ & \Planck2013 & 0 & $3.55$ & $-0.066$ & $1.85$ & $0.42$ & $13.78$ & $-0.16$ \\
\hline
\bahamas & \textsc{agn} & $7.8$ & \WMAP9 & 0.06 & $3.41$ & $-0.065$ & $2.09$ & $0.41$ & $13.84$ & $-0.13$ \\
\bahamas & \textsc{agn} & $7.8$ & \WMAP9 & 0.12 & $3.42$ & $-0.071$ & $2.09$ & $0.42$ & $13.88$ & $-0.16$ \\
\bahamas & \textsc{agn} & $7.8$ & \WMAP9 & 0.24 & $3.36$ & $-0.069$ & $2.30$ & $0.40$ & $13.86$ & $-0.14$ \\
\bahamas & \textsc{agn} & $7.8$ & \WMAP9 & 0.48 & $3.28$ & $-0.071$ & $2.65$ & $0.39$ & $13.86$ & $-0.10$ \\
\hline
\bahamas & \textsc{agn} & $7.8$ & \Planck2015 & 0.06 & $3.45$ & $-0.057$ & $2.03$ & $0.40$ & $13.82$ & $-0.13$ \\
\bahamas & \textsc{agn} & $7.8$ & \Planck2015 & 0.12 & $3.44$ & $-0.058$ & $2.07$ & $0.40$ & $13.83$ & $-0.14$ \\
\bahamas & \textsc{agn} & $7.8$ & \Planck2015 & 0.24 & $3.41$ & $-0.058$ & $2.20$ & $0.40$ & $13.84$ & $-0.12$ \\
\bahamas & \textsc{agn} & $7.8$ & \Planck2015 & 0.48 & $3.37$ & $-0.065$ & $2.54$ & $0.40$ & $13.84$ & $-0.10$ \\
\hline
\cosmoowls & \textsc{agn} & $8.0$ & \WMAP7 & 0 & $3.13$ & $-0.046$ & $2.26$ & $0.40$ & $13.56$ & $-0.09$ \\
\cosmoowls & \textsc{agn} & $8.5$ & \WMAP7 & 0 & $3.19$ & $-0.055$ & $2.03$ & $0.40$ & $14.28$ & $-0.11$ \\
\cosmoowls & \textsc{agn} & $8.7$ & \WMAP7 & 0 & $3.22$ & $-0.057$ & $1.77$ & $0.40$ & $14.83$ & $0.57$ \\
\hline
\cosmoowls & \textsc{agn} & $8.0$ & \Planck2013 & 0 & $3.23$ & $-0.039$ & $2.07$ & $0.40$ & $13.53$ & $-0.09$ \\
\cosmoowls & \textsc{agn} & $8.5$ & \Planck2013 & 0 & $3.30$ & $-0.046$ & $1.88$ & $0.40$ & $14.26$ & $-0.13$ \\
\cosmoowls & \textsc{agn} & $8.7$ & \Planck2013 & 0 & $3.38$ & $-0.056$ & $1.58$ & $0.42$ & $14.79$ & $0.29$ \\
\hline
\cosmoowls & \textsc{no-cool} & - &\WMAP7 & 0 & $4.22$ & $0.015$ & 0.00 & 0.00 & $12.39$ & $-0.15$ \\
\cosmoowls & \textsc{ref} & -& \WMAP7 & 0 & $3.79$ & $-0.007$ & $3.92$ & $0.27$ & $13.20$ & $-0.41$ \\
\hline
\end{tabular}
\label{tab:baryon_parameters}
\end{center}
\end{table*}

The best-fitting baryon feedback parameters for all the simulations we considered are listed in Table~\ref{tab:baryon_parameters}. All fitted models have an RMS error of less than one per cent. We mainly focussed on simulations that include realistic AGN feedback, but we also constrained our model using the \cosmoowls \textsc{ref} and \textsc{no-cool} simulations, which are not considered to be realistic. However, the fact that the model is still able to fit these simulations demonstrates some level of robustness. We see some obvious trends in our best-fit parameters, particularly with the AGN `strength', which is governed in the simulations by the sub-grid heating parameter $T_\mathrm{AGN}$ -- this is not a physical parameter that could be measured in the Universe. As feedback strength is increased $M_{\mathrm{b}}$ increases, which makes physical sense as more violent feedback expels more gas from the haloes. We also see a decrease in the star fraction, $f_*$, as $T_\mathrm{AGN}$ increases, which indicates that star formation is being suppressed by AGN feedback. Surprisingly, we see a preference for $f_*$ to increase with $z$ in all AGN feedback models, which taken literally would suggest that halo star fractions were higher in the past. This might be possible, if star formation peaked at higher redshifts and then shut off. However, this trend was not seen in \cite{Mead2020a}, which suggests that the relatively simplicity of the modelling presented here means that $f_*$ is capturing physics not associated directly with stars. In all simulations with AGN we see a preference for a decrease in the concentration-mass parameter $B$ away from the standard value of $4$. However, in \bahamas, $B$ decreases with feedback strength while in \cosmoowls $B$ increases. This difference may be due to the different hydrodynamical implementations between the two simulation suites, or it may point to a more complicated relationship between feedback strength and the back-reaction effect on the halo concentration. It may also be due to our model being overly simplistic since we take $B$ and $f_*$ to be independent of halo mass or because we do not attempt to explicitly model the bound or ejected gas profiles. In general, the feedback in \cosmoowls has a stronger impact on the power spectrum than that in \bahamas, and it is possible that there is a non-monotonic relationship between the feedback strength and the effect on $B$. As shown in \cite{McCarthy2018}, the power spectrum response to feedback when the neutrino mass is varied is quite minimal. This is reflected in the relatively similar best-fitting model parameters, particularly for the \Planck2015 case where the other cosmological parameters are all adjusted to be best-fitting for the \Planck data, which preserves the linear spectrum shape and therefore the feedback response. 

\subsection{Single-parameter model}

\begin{figure*}
\begin{center}
\includegraphics[height=18.5cm,angle=270]{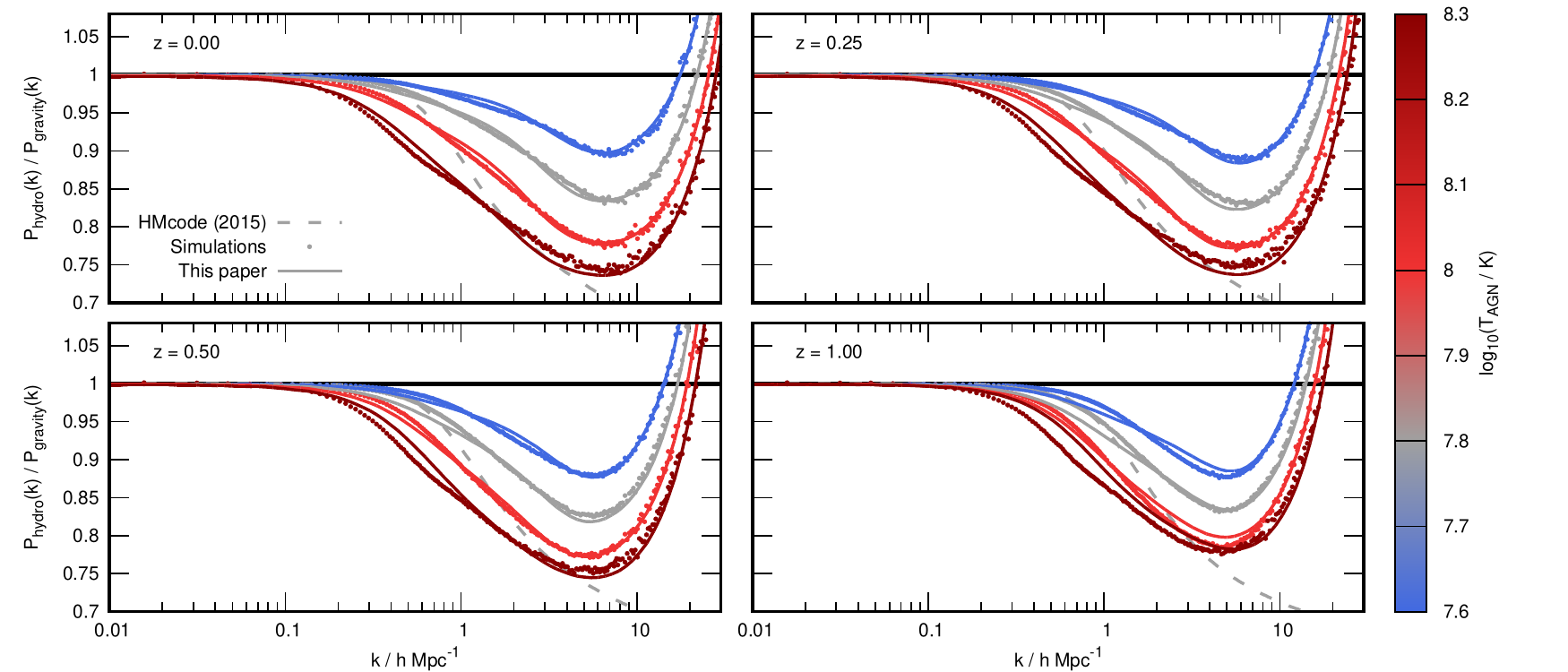}
\end{center}
\caption{Comparison of our single-parameter baryonic feedback response model against three different versions of \bahamas AGN feedback simulations with the \WMAP9 cosmology that differ only by their values of the `sub-grid heating' temperature (blue, grey, and light-red for $\log_{10}(T_\mathrm{AGN}/\mathrm{K})=7.6$, $7.8$, and $8.0$). The dots show the simulation measurements while the lines show the \hmcode model. In all cases we see that baryonic feedback suppresses power, starting at $k\sim0.1\iMpc$ with a maximum suppression effect of tens of per cent at $k\sim7\iMpc$, which is followed by a sharp rise in power. The suppression is caused by AGN feedback expelling gas from haloes while the rise in power is caused by galaxy formation in halo centres. The dark-red curve shows the \cosmoowls extreme AGN simulation with the \Planck2013 cosmology. This feedback scenario is quite well matched with the same single-parameter feedback model with an increased effective AGN temperature of $\log_{10}(T_\mathrm{AGN}/\mathrm{K})=8.3$. The dashed-grey line shows the AGN feedback model from the previous versions of \hmcode, which suppresses the power more than any of the scenarios shown here for $k>5\iMpc$ and the onset of the suppression is at slightly smaller scales compared to the \bahamas and \cosmoowls simulations.}
\label{fig:feedback_model}
\end{figure*}

\begin{table}
\begin{center}
\caption{Values for our single-parameter baryonic feedback model as a function of the sub-grid heating temperature in the \bahamas feedback models. In the formulae below $\theta=\log_{10}(T_\mathrm{AGN}/10^{7.8}\Kelvin)$. In equation~(\ref{eq:gas_fraction}) $\beta=2$ has been fixed. Parameter $X(z)$ is constructed from $X_0$ and $X_z$ as $X(z) = X_0\times 10^{z X_z}$.} 
\begin{tabular}{c c c}
\hline
Parameter & Equation & \bahamas formula \\
\hline
$B_0$ & \ref{eq:concentration_mass} & $3.44-0.496\theta$ \\
$B_z$ & \ref{eq:concentration_mass} & $-0.0671-0.0371\theta$ \\
$f_{*,0}/10^{-2}$ & \ref{eq:feedback_model} & $2.01-0.30\theta$ \\
$f_{*,z}$ & \ref{eq:feedback_model} & $0.409+0.0224\theta$ \\
$\log_{10}(M_{\mathrm{b},0}/\Msun)$ & \ref{eq:gas_fraction} & $13.87+1.81\theta$ \\
$M_{\mathrm{b},z}$ & \ref{eq:gas_fraction} & $-0.108+0.195\theta$ \\
\hline
\end{tabular}
\label{tab:feedback_model}
\end{center}
\end{table}

It may be preferable to have a less general, but single-parameter, model for baryonic feedback. Since degeneracies may exist between the $6$ parameters of our model, top-hat priors may not represent the true uncertainty, while also slowing down MCMC analyses. Based on the \bahamas best-fitting parameters in Table~\ref{tab:baryon_parameters} we see roughly linear trends between parameters and the feedback temperature. We therefore propose the model detailed in Table~\ref{tab:feedback_model}, where all $6$ feedback parameters have been linearly fitted as functions of  $\log_{10}(T_\mathrm{AGN}/\mathrm{K})$ for the \bahamas simulations. Given our modelling it is not possible to have a single-parameter model that works for both \cosmoowls and \bahamas since we observe the opposite trend in the mass-concentration parameter $B$ with $T_\mathrm{AGN}$ in each case. We therefore present a model only for \bahamas, although it would be straightforward to develop a similar model for \cosmoowls. In Fig.~\ref{fig:feedback_model} we show the performance of our single-parameter model against the three different $T_\mathrm{AGN}$ \bahamas simulations to which it was fitted. We see that the single-parameter model response follows the simulation response to within a couple of per cent across the entire range of scales shown. The largest errors are at the largest scales at $z=1$, where the scale-dependence of the departure of the response from unity in the model does not follow the simulations particularly well. However, the amplitude of the maximum dip in power and the smaller-scale increase in power are both well recovered. It should be noted that this one-parameter model has no limit in which the `gravity only' result is recovered. For comparison, in Fig.~\ref{fig:feedback_model} we also show the feedback model from the previous versions of \hmcode. The suppression in the previous model activates at smaller scales compared to the new model, predicts more suppression than all scenarios discussed in this paper for $k>5\iMpc$, and has no upturn because there was no attempt to model star formation in \cite{Mead2015b}. 


In Fig.~\ref{fig:feedback_model} we also show the performance of the same single-parameter model against the extreme \textsc{agn} $8.7$ simulation from the \cosmoowls suite. This feedback model was not used in the generation of our feedback model, but is well matched when assigned with an effective \bahamas $\log_{10}(T_\mathrm{AGN}/\mathrm{K})=8.3$. The other \cosmoowls \textsc{agn} $8.0$ and $8.5$ models can be similarly well matched with effective $\log_{10}(T_\mathrm{AGN}/\mathrm{K})=7.7$ and $8.0$ respectively. We should not be surprised that the effective \bahamas AGN temperature assigned to these \cosmoowls feedback scenarios is different from their \cosmoowls temperature because the underlying physical feedback model differs between the two simulation suites. However, it is pleasing that the same physical model can be made to work for both suites. Of all the feedback models the extreme \cosmoowls model shown in the Figure is the least well matched, with the largest errors at $z=1$ for $k < 3 \iMpc$. 


Our single-parameter model is calibrated on the \bahamas \WMAP9 \citep{Hinshaw2013} simulations, therefore we should compare it to the alternative \bahamas with \Planck \citeyear{Planck2013XVI} cosmology simulation. For the same sub-grid heating temperature, the simulations show that the \Planck2013 cosmology has less feedback-induced suppression, with a maximum of $\simeq 13$ per cent at $k\simeq 7\iMpc$ at $z=0$, compared to $\simeq 17$ per cent with the \WMAP9 cosmology. If we evaluate our single-parameter feedback model for the \Planck cosmology, the model \emph{does} predict less suppression, but with the maximum reduced from $\simeq 17$ to only $\simeq 15$, rather than as far as $\simeq 13$ per cent. Within our model, this change at fixed $T_\mathrm{AGN}$ arises due to the different baryon fractions\footnote{The baryon-mass fraction, $\Ob/\Om$, in the \WMAP9 cosmology is $\simeq0.166$ while it is $\simeq0.154$ in \Planck2013.} in each cosmology as because of differences in the halo mass functions. It is encouraging that this change with cosmology predicted by the model is in the correct direction, but disheartening that it is not quite strong enough. Table~\ref{tab:baryon_parameters} tells us that the \Planck cosmology prefers a slightly lower value of $M_{\mathrm{b},0}$ compared to \WMAP~9 ($\log_{10}(M_{\mathrm{b},0}/\Msun)=13.78$, rather than $13.84$). We tried to change the parameterisation in equation~(\ref{eq:gas_fraction}) for $M_\mathrm{b}$ to be proportional to the non-linear mass, $\sigma(M_*,z)=\delta_\mathrm{c}(z)$, but this trends in the wrong direction\footnote{At $z=0$ in \WMAP9 $M_*=10^{12.61}\Msun$ whereas in \Planck2013 $M_*=10^{12.74}\Msun$}. The simulations demonstrate that cosmologies with higher baryon fractions have a larger feedback-induced suppression in the matter power spectrum, presumably because a larger fraction of the halo mass is expelled. The dependence on the baryon fraction is built into our model via equation~(\ref{eq:feedback_model}) but this comparison suggests that the true dependence is slightly different. In any case, we find that the suppression in the \Planck cosmology is well fitted, with an RMS error below one per cent, by our single-parameter model but with a slightly reduced $\log_{10}(T_\mathrm{AGN}/\mathrm{K})=7.75$.

The \cite{vanDaalen2020} library contains cosmologies that combine baryonic feedback with massive neutrinos. There are two cases to consider: \WMAP9, where as the neutrino mass is increased $\sigma_8$ is decreased and \Planck2015, where as the neutrino mass is increased all other cosmological parameters are changed to maintain a good fit to the \Planck data. In the \WMAP9 case there is a weak dependence on the suppression with neutrino mass, with the trend that the suppression slightly increases as the neutrino mass increases. If we evaluate our single-parameter model we find the correct direction for the trend, but one that is slightly stronger than that seen in the simulations. However, since the change seen in the simulations is very slight, the eventual model error is quite small. In the extreme $M_\nu=0.48\eV$ case, the difference in suppression between this and the $M_\nu=0$ case is predicted to be around twice as large as is seen in the data, but the maximum error is only $2$ per cent -- an improved match can be made with a slight increase in the corresponding AGN temperature. The same holds for the \Planck2015 massive-neutrino cases, but the trends, and the eventual error in our model, are both smaller.

\subsection{Priors}

\begin{table*}
\begin{center}
\caption{Recommended priors for our baryonic feedback model that are able to encompass specific hydrodynamic feedback scenarios. We list parameters that encompass the \bahamas suite, those that encompass both \bahamas and \cosmoowls, and those that additionally include the gravity-only case, which would be correct if there were no feedback. The first column lists the scenario, the next $6$ columns list the $6$ parameters of our feedback model, while the two values in each column indicate the lower and upper bound. The final column lists the $T_\mathrm{AGN}$ values that should be used for our single-parameter model. Note that our single-parameter model does not encompass the gravity only scenario.} 
\begin{tabular}{c c c c c c c c c c c c c c c}
\hline
Scenario &  \multicolumn{2}{c}{$B_0$} & \multicolumn{2}{c}{$B_z$} & \multicolumn{2}{c}{$f_{*,0}/10^{-2}$} & \multicolumn{2}{c}{$f_{*,z}$} & \multicolumn{2}{c}{$\log_{10}(M_{\mathrm{b},0}/\Msun)$} & \multicolumn{2}{c}{$M_{\mathrm{b},z}$} & \multicolumn{2}{c}{$\log_{10}(T_\mathrm{AGN}/\mathrm{K})$} \\
\hline
\bahamas & $3.28$ & $3.55$ & $-0.075$ & $-0.057$ & $1.85$ & $2.65$ & $0.39$ & $0.42$ & $13.52$ & $14.24$ & $-0.16$ & $-0.05$ & $7.6$ & $8.0$ \\
 + \cosmoowls & $3.13$ & $3.55$ & $-0.075$ & $-0.039$ & $1.58$ & $2.65$ & $0.39$ & $0.42$ & $13.52$ & $14.83$ & $-0.16$ & $0.57$ & $7.6$ & $8.3$ \\
 + gravity-only & $3.13$ & $4.00$ & $-0.075$ & $-0.039$ & $0$ & $2.65$ & $0.39$ & $0.42$ & $0$ & $14.83$ & $-0.16$ & $0.57$ & \multicolumn{2}{c}{-} \\
\hline
\end{tabular}
\label{tab:priors}
\end{center}
\end{table*}

Based on the data in Table~\ref{tab:baryon_parameters} reasonable priors could be adopted for our $6$ parameter model which would cover the full range of feedback scenarios from \cite{vanDaalen2020}. These ranges could be tightened if the more extreme, and less physically reasonable, \cosmoowls simulations were discarded. The gravity-only model is recovered in the limit $B\to4$, $M_\mathrm{b}\to0$ and $f_*\to0$. In Table~\ref{tab:priors} we explicitly state our recommended prior ranges for each scenario. We also include a prior range for our single-parameter model. In the \bahamas case we recommend varying between the range of $T_\mathrm{AGN}$ probed by the \bahamas simulations: $7.6<\log_{10}(T_\mathrm{AGN}/\Kelvin)<8.0$. This temperature range was chosen so that the \bahamas simulations reproduced the full range of halo-gas fraction observations \citep{McCarthy2017}, which \cite{Debackere2020} and \cite{vanDaalen2020} have shown is closely linked to the eventual magnitude of the feedback suppression. If one wanted to consider more extreme feedback scenarios, then in the \bahamas + \cosmoowls case we recommend extending the upper bound such that $7.6<\log_{10}(T_\mathrm{AGN}/\Kelvin)<8.3$ is explored, since this encompasses the full range of \bahamas and \cosmoowls feedback, as demonstrated in Fig.~\ref{fig:feedback_model}.

\section{Summary and discussion}
\label{sec:summary}

We have presented \hmcodett, an updated and augmented halo model for the non-linear matter--matter power spectrum that has an RMS error compared to simulated data of less than $2.5$ per cent across a wide range of cosmologies, scales and redshifts. This represents a significant improvement over previous \halofit and \hmcode versions. The span of cosmological parameters also ensures that this encompasses a wide range of linear spectrum shapes, so that we can be confident that our model is capturing many of the non-trivial aspects of non-linear evolution. Our augmented halo model is parametrized in terms of the linear power spectrum, transformed into the mass function and concentration parameters via the halo model, as well as in terms of variables that pertain to the linear spectrum properties such as $\sigma_8(z)$ or $n^\mathrm{eff}(z)$. We believe that parameterising the model this way is advantageous because there is no specific dependence on the standard cosmological parameters so that the model may be applied to novel cosmological scenarios as and when they are invented. Parameters in \hmcodett were fitted to simulated data taken from the node cosmologies of \frankenemu. \hmcodett was also compared to \miratitan, to which it was not fitted, and was shown to out perform competing models. It is possible that performance would be further improved if we had additionally fitted to the node cosmologies of \miratitan, but we have not done this and instead prefer to leave \miratitan as a fair test. \hmcodett is also rapid; the non-linear power for $128$ $k$ values, at $n_z=16$ redshifts for a single cosmology can be evaluated in $160\,\mathrm{ms}$ on a single core of a modern laptop, with the compute time scaling as $t\simeq(22+5n_z)\,\mathrm{ms}$. The computation is slower for massive-neutrino cosmologies, scaling as $t\simeq(1500+9n_z)\,\mathrm{ms}$. These timings are similar to previous versions of \hmcode.

Section~\ref{sec:ingredients} provides a comparison of the accuracy of vanilla halo model predictions compared to simulations -- both in terms of the mean prediction across a range of cosmologies as well as the variance about this mean. This allowed us to select an optimal set of ingredients for \hmcodett. We demonstrated that halo-model predictions when using haloes defined using the $200$-times critical density definition were particularly poor compared to other halo definitions. We also demonstrated that predictions using the \cite{Sheth1999} mass function were better compared to when using \cite{Tinker2010}, and we suggest that this because the mass function of \cite{Sheth1999} was fitted to a broader range of cosmologies. We are not aware of any similar analyses, and this conclusion may be useful when decisions are taken regarding ingredients to use in other halo model calculations, such as for the galaxy power spectrum or for the tSZ effect. 

Perturbative effects are missing from the standard halo model calculation. In \hmcode they are included in an approximate way via an ad-hoc damping term applied to the linear power in the two-halo term, as well as via the damping of BAO amplitude via a well-known perturbation theory result. We have yet to include actual perturbation theory calculations within \hmcode but this would be a fruitful direction for future research, as long as the speed of the calculation could be kept under control. Combing perturbation theory and the halo model has been considered by other authors including \cite{Valageas2011}, \cite{Mohammed2014b}, \cite{Seljak2015}, \cite{Hand2017}, \cite{Cataneo2019} and \cite{Philcox2020}. While these results are promising, the authors do not achieve the same accuracy across the range of scales, redshifts and cosmologies as has been achieved using \hmcode, although most achieve better accuracy at larger scales. In these approaches the one-halo term is often thrown away, or treated as a generic series expansion. We suggest that keeping the cosmological information in the one-halo term, which enters via the mass function and halo profile, is preferable since we have seen that the ability of the halo model to track the power in the non-linear regime can be greatly improved with the correct choice of these ingredients.

We use the \cite{Bullock2001} halo concentration--mass relation in \hmcode. While this works well for our purposes, as demonstrated in Section~\ref{sec:ingredients}, it may be preferable to update the relations to those presented in \cite{Correa2015}, \cite{Ludlow2016} or \cite{Diemer2019}, where the concentration is tied to the halo mass-accretion history. However, each of these relations is calibrated for halo mass definitions of $200$ times the critical density, whereas we demonstrated that our halo-model predictions were far superior when using the virial halo definition. It is tempting to try to change the halo definition using the well-known method involving a fixed NFW profile and scale radius, but we do not do this here as it is not obvious that this method produces the haloes with the same characteristics as would be observed if a halo finder with a different over-density threshold were used to identify haloes in the same simulation.

In order to reproduce accurate power spectra we were forced to introduce a number of free-parameters and ad-hoc changes to the standard halo model. It is tempting to view these as indicative of some underlying fundamental model, but we caution against this. For example, we were forced to introduce the $\eta$ parameter, which bloats haloes as a function of mass, reaching a $\simeq15$ per cent size increase for haloes of $10^{15}\Msun$ at $z=0$. The unphysical bloating is necessary to get a good match to simulated spectra, but it is alarming that the haloes to which the model pertains are not the same as haloes in the real Universe. Another cause for concern is the $\alpha$ parameter, which smooths the transition between the two- and one-halo terms, and is absolutely essential in getting good matches to the simulated data. Could the same $\alpha$ work for other halo model calculations that have similar problems in the transition region? Alas, a simple thought experiment proves that this is not possible. Consider a field that is made up of the sum of two constituent fields (for example matter, being made of CDM and baryons). If the auto- and cross-power spectrum of each field separately can be written as a sum of two- and one-halo terms then it can only be that the total field can also be written as a sum of two- and one-halo terms if $\alpha=1$. Clearly a less crude technique than our ad-hoc smoothing will be required to solve this fundamental issue.

We have also presented a halo model for the response in the power spectrum to baryonic feedback. Simple recipes for gas expulsion and star formation are included, as well as the impact of feedback on the halo concentration. The feedback model was fitted to simulation data, taken from the library of \cite{vanDaalen2020}, with an RMS error of less than one per cent for all feedback models that we considered ($k<20\iMpc$; $z<1$). We provide the best-fitting parameters of our model to each simulation and suggest that these can be used to generate an informative prior range that may be sampled over in future weak-lensing data analyses to immunise cosmological constraints against the unknown magnitude of baryonic feedback. Alternatively, we provide a single-parameter feedback model, parametrized in terms of the sub-grid AGN heating temperature of the \bahamas simulations, which may be robust to modest extrapolation. We noted that this model did not capture the full cosmology dependence of the feedback correctly, underestimating the shift between the \WMAP9 and \Planck2013 cosmology by half, although still only making a maximum $2$ per cent error in the absolute response. However, the \Planck2013 feedback suppression \emph{was} well matched with a slightly reduced effective AGN temperature ($\log_{10}(T_\mathrm{AGN}/\mathrm{K})$ = 7.75, rather than 7.8). We consider this to be acceptable since $T_\mathrm{AGN}$ is not a physical parameter and would be marginalised over in future analyses, so that it is only important that physical feedback suppression is captured by some value of the $T_\mathrm{AGN}$ parameter and that we marginalise over a suitable range. The single-parameter model also slightly over predicts the change in response as massive neutrinos are introduced, but again only making small absolute errors because the neutrinos have a minimal effect on the suppression. Despite this, in both cases excellent results can still be obtained if one slightly modifies the AGN temperature. In its current form, the single-parameter model maps an AGN temperature to a unique value of $M_{\mathrm{b}}$, the mass below which a halo has lost half of its initial gas content, and this removes a fixed fraction of the baryons originally in a halo. Clearly this is simplistic, and cosmologies with different baryon fractions may be differently affected for the same AGN temperature. The cosmology dependence of the amplitude of feedback clearly warrants further research, and for this it would be useful to have access to hydrodynamic simulations over a wider range of cosmologies. We found it more difficult to fit our model to $z>1$, particularly considering the $z$ dependence we adopted for our feedback model. It is probably sufficient for current lensing studies to concentrate on the feedback only at the lowest redshifts, because it is a small-scale effect, but in the not-too-distant future the increased fidelity of observational data may force this to be revisited.

\section*{Data availability}
The code developed and used in this article is available at \hmcodelink. The data underlying this article will be shared upon request to the corresponding author.

\section*{Acknowledgements}
AJM and TT have received funding from the Horizon 2020 research and innovation programme of the European Union under Marie Sk\l{}odowska-Curie grant agreements No. 702971 and 797794 respectively. AJM, TT and CH acknowledge support from the European Research Council under grant number 647112. CH acknowledges support from the Max Planck Society and the Alexander von Humboldt Foundation in the framework of the Max Planck-Humboldt Research Award endowed by the Federal Ministry of Education and Research. The authors acknowledge useful conversations with Licia Verde, Simeon Bird, Sarah Brown, and Mijin Yoon.

\label{lastpage}

\footnotesize{
\setlength{\bibhang}{2.0em}
\setlength\labelwidth{0.0em}
\bibliographystyle{mnras}
\bibliography{meadbib}
}

\normalsize

\appendix

\section{Fitting functions for $\dc$ and $\Dv$}
\label{app:fitting_functions}

In this Appendix we detail fitting functions for the spherical-collapse parameters $\dc$ and $\Dv$ that are valid for homogeneous dark energy models that also work for cosmologies with substantial curvature. These fitting functions originally appeared in \cite{Mead2017}, but are repeated here with some minor alterations that make them appropriate for massive-neutrino cosmologies. An alternative \LCDM fitting function for $\dc$ can be found in \cite{Nakamura1997} and for $\Dv$ in \cite{Bryan1998}. For matter-dominated open and closed cosmologies exact formulae may be derived \citep[\eg][]{b:Coles}. The fitting functions presented here are constructed in such a way that they do not depend explicitly on any dark-energy parametrization, and are instead parametrized in terms of the matter density, linear growth and the integrated growth, which are defined below.

\begin{table}
\begin{center}
\caption{Best-fitting parameters for the $\dc$ and $\Dv$ fitting functions in equations~(\ref{eq:dc_fit}) and (\ref{eq:Dv_fit}). The $\alpha_i$ exponents are defined in these equations while the $p_{i,j}$ coefficients of the quadratics are defined in equation~(\ref{eq:quadratic}).}
\begin{tabular}{c c c c}
\hline
$\dc$ & Value & $\Dv$ & Value\\
\hline
$p_{1,0}$ & $-0.0069$ & $p_{3,0}$ & $-0.79$ \\
$p_{1,1}$ & $-0.0208$ & $p_{3,1}$ & $-10.17$\\
$p_{1,2}$ & $0.0312$ & $p_{3,2}$ & $2.51$\\
$p_{1,3}$ & $0.0021$ & $p_{3,3}$ & $6.51$\\
$\alpha_1$ & $1$ & $\alpha_3$ & $1$\\
$p_{2,0}$ & $0.0001$ & $p_{4,0}$ & $-1.89$\\
$p_{2,1}$ & $-0.0647$ & $p_{4,1}$ & $0.38$\\
$p_{2,2}$ & $-0.0417$ & $p_{4,2}$ & $18.8$\\
$p_{2,3}$ & $0.0646$ & $p_{4,3}$ & $-15.87$\\
$\alpha_2$  & $0$ & $\alpha_4$ & $2$ \\
\hline
\end{tabular}
\label{tab:fit_params}
\end{center}
\end{table}

We parametrize a fitting function for $\dc$ via
\begin{equation}
\frac{\dc}{1.686(1-0.041 f_\nu)}=1+\sum_{i=1,2} f_{i}\left(\frac{g}{a},\frac{G}{a}\right)\left[\log_{10}\Om(a)\right]^{\alpha_i}\ ,
\label{eq:dc_fit}
\end{equation}
and for $\Dv$ via
\begin{equation}
\frac{\Dv}{177.7(1+0.763 f_\nu)}=1+\sum_{i=3,4} f_{i}\left(\frac{g}{a},\frac{G}{a}\right)\left[\log_{10}\Om(a)\right]^{\alpha_i}\ ,
\label{eq:Dv_fit}
\end{equation}
with $f_i$ being the quadratics 
\begin{equation}
f_i (x,y)=p_{i,0}+p_{i,1}(1-x)+p_{i,2}(1-x)^2+p_{i,3}(1-y)\ .
\label{eq:quadratic}
\end{equation}
These functions ensure that the Einstein-de Sitter values\footnote{More correctly $\dc=(3/20)\times(12\pi)^{2/3}$ and $\Dv=18\pi^2$.} $\dc\simeq1.686$ and $\Dv\simeq178$ are recovered when $\Om(a)=1$. The best-fitting parameter values are given in Table~\ref{tab:fit_params}.

In these equations, $g(a)$ is the \emph{unnormalized} linear-theory growth function; the solution to
\begin{equation}
g''+\left(\frac{A}{H^2}+2\right)\frac{g'}{a}=\frac{3}{2}\Omega_\mathrm{m}(a)\frac{g}{a^2}\ ,
\label{eq:growth}
\end{equation}
where the dashes represent derivatives with respect to $a$, and taking the initial condition $g=a$ when $a\ll 1$, which is the Einstein-de Sitter growing mode solution that is valid at early times when radiation is neglected. $A\equiv\ddot{a}/a$ and $H=\dot{a}/a$ is the standard Hubble parameter. We also define the integrated growth via
\begin{equation}
G(a)=\int_0^a\frac{g(\tilde{a})}{\tilde{a}}\,\mathrm{d}\tilde{a}\ ,
\label{eq:integrated_growth}
\end{equation}
where the tilde denotes a dummy variable. For the Einstein-de Sitter growing mode $G(a)=a$. For reference, for $\Om=0.3$ \LCDM $g(a=1)\simeq0.779$ and $G(a=1)\simeq0.930$.

The spherical-collapse calculations to which our relations were fitted do not include the buoying effect of dark-energy in the calculation of the final virial density of haloes, and so consequently our $\Dv$ does not map exactly to that of the \cite{Bryan1998} formula. For example, for a standard $\Om=0.3$ \LCDM model the \cite{Bryan1998} relation gives $\Dv\simeq337$, whereas the \cite{Mead2017} relation gives $\Dv\simeq308$. Attempts to include general dark energy in calculations of the virial density can be found in \cite{Mota2004} or \cite{Wintergerst2010}, but are not included here.

As shown in \cite{Mead2016} these spherical-collapse results can be applied to massive neutrino cosmologies by making the multiplicative correction that appears in the denominators in the left-hand side of equations~(\ref{eq:dc_fit}) and (\ref{eq:Dv_fit}) compared to the results presented in \cite{Mead2017}. These values were calculated for matter-dominated models, but the same correction also works for massive-neutrino cosmologies with exotic dark energy equations of state. In the massive neutrino case $g(a)$, $G(a)$ and $\Omega_\mathrm{m}(a)$ that appear in the fitting functions are all evaluated in an equivalent massless neutrino cosmology with the mass in neutrinos converted to CDM.

\section{halofit and massive neutrinos}
\label{app:halofit_neutrinos}

\halofit has three separate versions all published by different groups. The first, \cite{Smith2003}, version is inaccurate compared to modern simulations, as can be seen in Fig.~\ref{fig:FrankenEmu}, a consequence of it having been fitted to low-resolution simulations that systematically under-predict power at small scales. Chronologically, the \cite{Bird2012} and the \cite{Takahashi2012} papers were published at similar times, and both provided updates of the original \cite{Smith2003} \halofit fitting function. \cite{Takahashi2012} improved the model to be generally more accurate by fitting to more modern simulations and also introduced an explicit dependence on $w$, the dark-energy equation of state. In doing so, all parameters in the model were refitted to a modern simulation suite. In contrast, the focus of \cite{Bird2012} was on cosmologies with massive neutrinos, which were ignored by \cite{Takahashi2012}. In \cite{Bird2012} most parameters of the original \cite{Smith2003} version were left unaltered and terms were added with explicit dependencies on the neutrino mass. Because the base model was left almost unaltered, the overall accuracy of \cite{Bird2012} version for standard cosmologies with no massive neutrinos was worse than that of \cite{Takahashi2012}. To rectify this, the \cite{Bird2012} neutrino model was later reconfigured using the more accurate \cite{Takahashi2012} \halofit as a base (Bird, private communication). This reconfiguration has not been published, but it is the default \halofit version in \camb. If the neutrino mass is zero then this `\camb' version of \halofit is exactly the \cite{Takahashi2012} version because the only new parameters vanish in this limit.

\begin{figure*}
\begin{center}
\hspace*{-0.15cm}\includegraphics[height=18cm,angle=270]{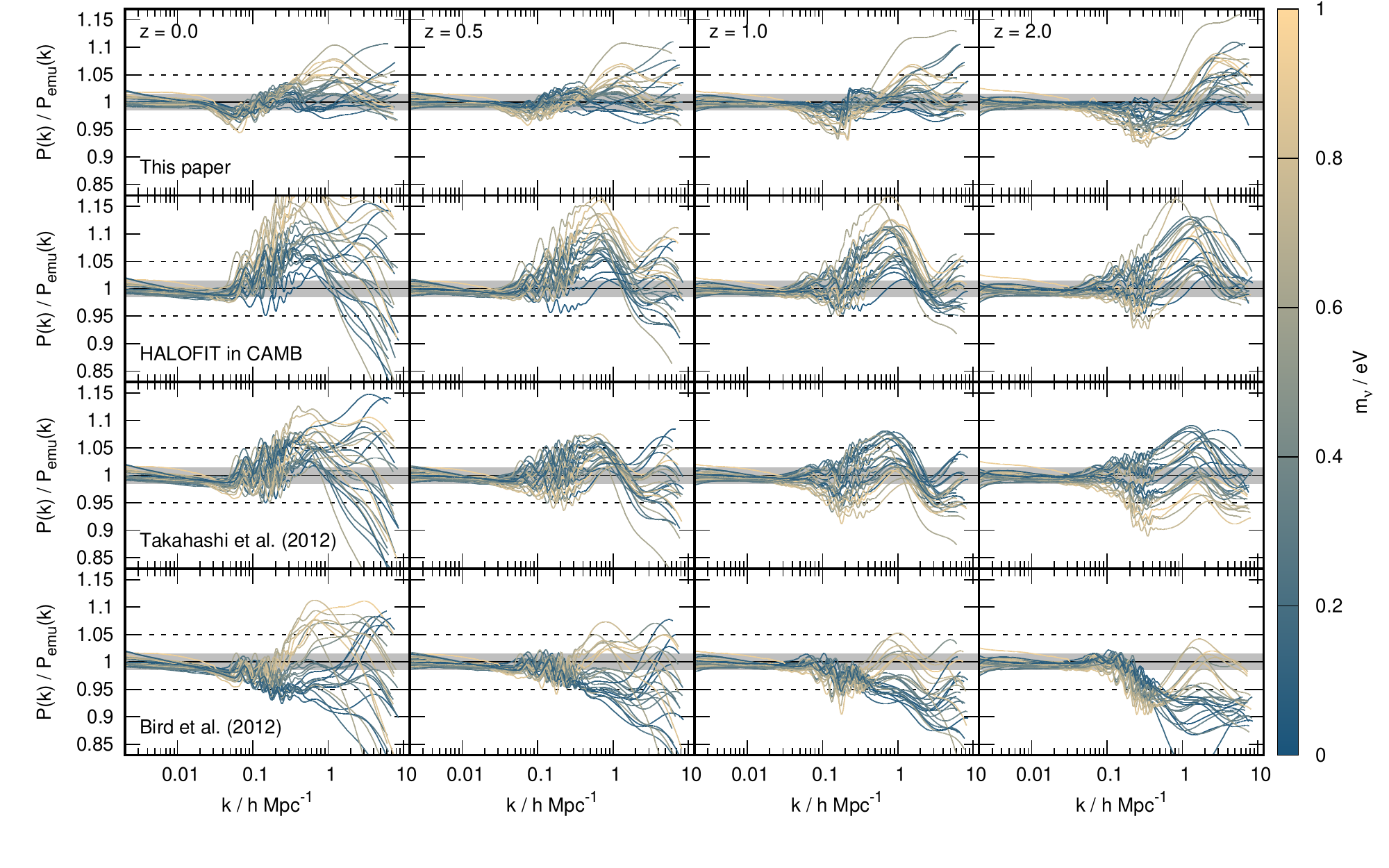}
\end{center}
\caption{Performance of different versions of \halofit compared to the massive-neutrino node cosmologies of \miratitan. Each line is a different cosmological model and these are colour-coded according to the neutrino mass. We show three versions of \halofit: the bottom row shows the original \protect\cite{Bird2012} version, which itself is a correction to the original of \protect\cite{Smith2003}; the third row shows the original of \protect\cite{Takahashi2012} which has no explicit dependence on neutrino mass; the second row shows an unpublished version of the \protect\cite{Bird2012} neutrino \halofit, which is the default in \camb; the top row shows \hmcodett for comparison. We note that the unpublished \halofit that contains explicit neutrino-mass dependence performs worse, especially for high neutrino masses, compared to the \protect\cite{Takahashi2012} version with no explicit neutrino mass dependence.}
\label{fig:halofit_neutrinos}
\end{figure*}

In Fig.~\ref{fig:halofit_neutrinos} we show the performance of different versions of \halofit compared to the node cosmologies of \miratitan that have massive neutrinos (26 of 36 nodes): We show the original \cite{Bird2012} model; the original version of \cite{Takahashi2012}, which has no explicit dependence on neutrino mass; and we show the unpublished version of the \cite{Bird2012} neutrino \halofit, which is the default model that appears in \camb. It is surprising to see that the \camb version, which is explicitly calibrated on cosmologies with massive neutrinos, performs worse than \cite{Takahashi2012} version, despite the latter having no explicit dependence on neutrino mass. This poor performance is especially striking for the highest neutrino masses. While there are some obvious systematic errors in the mean prediction of the \cite{Bird2012} version, it is interesting that the variance between the predictions for the different cosmologies is better than for the other two models around $k\sim0.3\iMpc$ and also that the variance is not significantly worse than for the other models at smaller scales.

It is probable that this can be explained by the different techniques used to include massive neutrinos in cosmological simulations. The three main techniques, in order of complexity, are:
\begin{itemize}
\item Linear neutrinos: are treated entirely via linear theory \citep[\eg][]{Agarwal2011}. Their energy density is included in the evolution of the background, and their contribution to the gravitational potential is included and affects the simulation particles, but there is no back-reaction on the neutrinos from the simulation particles.
\item Semi-linear neutrinos: have linear equations of motion for the neutrinos evolved either on a grid  \citep[][]{Brandbyge2009} or by solving the Boltzmann equation linearized with respect to the neutrino overdensity \citep{Ali-Haimoud2013} while the simulation is run. While these equations are linear, they are sourced by the non-linear gravitational potential that arises from the simulation particles so they do not evolve as in linear theory. Simulation particles know about the neutrino grid via their contribution to the gravitational potential.
\item Particle neutrinos: neutrinos are explicitly included as a separate particle species in the simulation, with different initial conditions compared to CDM and with initial thermal velocities \citep[\eg][]{Viel2010}.
\end{itemize}
While the final technique seems the most comprehensive there are pitfalls that must be avoided: one must decide on when to inject the neutrino particles in to the simulation, too early and the neutrinos are too hot and make circuits of the simulation volume, too late and non-linear evolution in the neutrino density is missed. More recently hybrid techniques have been developed to address this \citep[\eg][]{Bird2018}. 

It has been shown that simulations using the three different techniques give very similar results for the eventual non-linear matter power spectrum for low neutrino masses ($m_\nu \ltsim 0.3\eV$). This makes sense, since these neutrinos are the hottest and their evolution closely follows linear theory, even at late times, and they also contribute less to the energy budget of the cosmos due to their low mass (recall that $\Omega_\nu\propto m_\nu$). However, at high masses the results diverge. The \miratitan simulations of \cite{Lawrence2017} use the linear neutrinos and the simulations to which the \cite{Bird2012} \halofit model was fitted use particle neutrinos. In this paper, we are not in a position to comment on which of these options is better, but we suggest that differences between these simulation techniques are probably responsible for the fact that the \cite{Bird2012} additions seem to degrade the performance of \cite{Takahashi2012} for massive neutrino cosmologies when comparing to \miratitan. Note carefully that \hmcodett reproduces the results of the \miratitan emulator well, even at high neutrino masses. If it turned out that the \miratitan neutrino simulation technique was inaccurate at high neutrino masses, then so would be \hmcodett.

\section{Early dark energy}
\label{app:early_dark_energy}

\begin{figure*}
\begin{center}
\hspace*{-0.15cm}\includegraphics[height=18cm,angle=270]{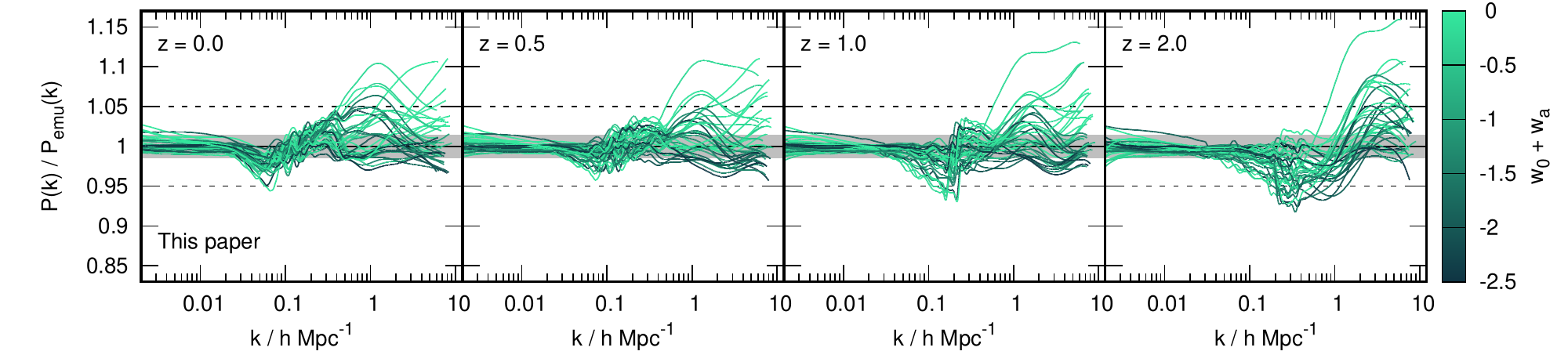}
\end{center}
\caption{The performance of the new version of \hmcode presented in this paper compared against the \miratitan node cosmologies, colour coded according to $w_0+w_a=w(a\to0)$. The results presented here are the same as in Fig.~\ref{fig:MiraTitan} but with a different colour bar. It is clear that the error in \hmcode for $k>0.5\iMpc$ is a strong function of the early-time value of the dark energy equation of state. Cosmologies with $w_0+w_a$ values close to zero can be considered `early dark energy' models, where the cosmological $\Omega_w(a\to0)$ can still be considerable.}
\label{fig:early_dark_energy}
\end{figure*}

Time-varying dark energy is often parametrized via the equation of state
\begin{equation}
w(a)=w_0+(1-a)w_a\ ,
\end{equation}
where $w_0$ is the value today and $w_0+w_a$ is the value at early times. In a standard \LCDM model the dark energy density decays very quickly as a function of redshift, becoming completely negligible at early times. In contrast, cosmologies where $w_0+w_a\simeq 0$ can have relatively high dark energy densities at early times, since the dark energy density essentially tracks the matter density. These cosmologies can even have appreciable dark energy densities around recombination, and are therefore known as `early dark energy' cosmologies. In Fig.~\ref{fig:early_dark_energy} we show the performance of \hmcodett against the \miratitan node cosmologies where the cosmologies have been colour coded according to the value of $w_0+w_a$, these are the same results as in Fig.~\ref{fig:MiraTitan} apart from the colour change. We see a strong correlation between the error made by \hmcodett and the early-time behaviour of the dark energy equation of state, with the largest errors made for cosmologies with the early-time equation of state close to zero. This could conceivably be because these cosmologies have the least standard structure formation pathway, given that dark energy is present at the universe even when the first haloes are forming. This is in contrast to say a model with $w_0=-1.3$ and $w_a=0$, because in this case, even though the equation of state is `strange' the dark energy only has an impact upon structure formation at very late times (with this becoming ever later as $w_0$ becomes more negative). It has been known since \cite{Navarro1997} that there is a strong correlation between the internal properties of haloes and their formation times. We suspect that the errors made by \hmcodett reflect that the halo concentrations in these cosmologies are not being captured by our combination of the \cite{Bullock2001} relation with the \cite{Dolag2004} dark-energy correction. However, Fig.~\ref{fig:early_dark_energy} shows that the error rarely exceeds $10$ per cent for $k>1\iMpc$ and only certain early-dark energy cosmologies seem to be badly affected.

\section{Emulator comparison}
\label{app:emulators}

\begin{table*}
\begin{center}
\caption{Cosmological parameter ranges covered by those power-spectrum emulators discussed in this Appendix. The table is split into parameters for: standard matter density (top), neutrinos, power spectrum normalisation, power spectrum shape, $h$, and dark energy (bottom). Note that $\Omega_\mathrm{cold}=\Omega_\mathrm{m}-\Omega_\nu=\Omega_\mathrm{c}+\Omega_\mathrm{b}$ and is different from $\Omega_\mathrm{c}$, which is the CDM density parameter. Only \miratitan, \euclid and \bacco consider massive neutrinos. Only \halofit considers the running of the spectral index $n_\mathrm{run}$.\cosmicemu and \frankenemu do not consider time-varying dark energy. All emulators assume a flat universe, so $\Omega_w=1-\Omega_\mathrm{m}$ always. With these considerations, the values of the `missing' cosmological parameters, marked with a -, can always be calculated from the given parameters, while those parameters marked with a * are not considered by an emulator (fixed to $0$ in all cases except $h$ for \cosmicemu, which is fixed via the well-measured angular extent of the CMB acoustic scale).}
\begin{tabular}{c c c c c c c c c c c c c}
\hline
\multirow{2}{*}{Parameter} &  \multicolumn{2}{c|}{\cosmicemu} &  \multicolumn{2}{c|}{\frankenemu} &  \multicolumn{2}{c|}{\miratitan} & \multicolumn{2}{c|}{\halofit} & \multicolumn{2}{c|}{\euclid} & \multicolumn{2}{c|}{\bacco} \\
\cline{2-13}
& min & max & min & max & min & max & min & max & min & max & min & max \\

\hline

$\Omega_\mathrm{m}$ & \multicolumn{2}{c}{-} & \multicolumn{2}{c}{-} & \multicolumn{2}{c}{-} & $0.21$ & $0.4$ & $0.24$ & $0.4$ & \multicolumn{2}{c}{-} \\

$\omega_\mathrm{m}$ & $0.12$ & $0.155$ & $0.12$ & $0.155$ & $0.12$ & $0.155$ & \multicolumn{2}{c}{-} & \multicolumn{2}{c}{-} & \multicolumn{2}{c}{-} \\

$\Omega_\mathrm{b}$ & \multicolumn{2}{c}{-} & \multicolumn{2}{c}{-} & \multicolumn{2}{c}{-} & \multicolumn{2}{c}{-} & $0.04$ & $0.06$ & $0.04$ & $0.06$\\

$\omega_\mathrm{b}$ & $0.0215$ & $0.0235$ & $0.0215$ & $0.0235$ & $0.0215$ & $0.0235$ & $0.02$ & $0.024$ & \multicolumn{2}{c}{-} & \multicolumn{2}{c}{-} \\

$\Omega_\mathrm{c}$ & \multicolumn{2}{c}{-} & \multicolumn{2}{c}{-} & \multicolumn{2}{c}{-} & \multicolumn{2}{c}{-} & \multicolumn{2}{c}{-} & \multicolumn{2}{c}{-} \\

$\omega_\mathrm{c}$  & \multicolumn{2}{c}{-} & \multicolumn{2}{c}{-} & \multicolumn{2}{c}{-} & $0.1$ & $0.13$ & \multicolumn{2}{c}{-} & \multicolumn{2}{c}{-} \\

\hline

$\Omega_\mathrm{cold}=\Omega_\mathrm{m}-\Omega_\nu$ & \multicolumn{2}{c}{-} & \multicolumn{2}{c}{-} & \multicolumn{2}{c}{-} & \multicolumn{2}{c}{-} & \multicolumn{2}{c}{-} & $0.23$ & $0.4$ \\


$\omega_\nu$ & \multicolumn{2}{c}{*} & \multicolumn{2}{c}{*} & $0$ & $0.01$ & \multicolumn{2}{c}{*} & \multicolumn{2}{c}{-} & \multicolumn{2}{c}{-} \\

$m_\nu/\eV$ & \multicolumn{2}{c}{*} & \multicolumn{2}{c}{*} & \multicolumn{2}{c}{-} & \multicolumn{2}{c}{*} & $0$ & $0.15$ & $0$ & $0.4$ \\

\hline

$\sigma_8$ & $0.61$ & $0.9$ & $0.616$ & $0.9$ & $0.7$ & $0.9$ & \multicolumn{2}{c}{-} & \multicolumn{2}{c}{-} & \multicolumn{2}{c}{-} \\

$\sigma_{8, \mathrm{cold}}$ & \multicolumn{2}{c}{-} & \multicolumn{2}{c}{-} & \multicolumn{2}{c}{-} & \multicolumn{2}{c}{-} & \multicolumn{2}{c}{-} & $0.73$ & $0.90$ \\

$A_\mathrm{s}/10^{-9}$ & \multicolumn{2}{c}{-} & \multicolumn{2}{c}{-} & \multicolumn{2}{c}{-} & $1.72$ & $2.58$ & $1.7$ & $2.5$ & \multicolumn{2}{c}{-} \\

\hline

$n_\mathrm{s}$ & $0.85$ & $1.05$ & $0.85$ & $1.05$ & $0.85$ & $1.05$ & $0.85$ & $1.05$ & $0.92$ & $1$ & $0.92$ & $1.01$ \\

$n_\mathrm{run}$ & \multicolumn{2}{c}{*} & \multicolumn{2}{c}{*} & \multicolumn{2}{c}{*} & $-0.2$ & $0.2$ & \multicolumn{2}{c}{*} & \multicolumn{2}{c}{*} \\

\hline

$h$ & \multicolumn{2}{c}{*} & $0.55$ & $0.85$ & $0.55$ & $0.85$ & \multicolumn{2}{c}{-} & $0.61$ & $0.73$ & $0.6$ & $0.8$ \\

\hline

$w_0$ & $-1.3$ & $-0.7$ & $-1.3$ & $-0.7$ & $-1.3$ & $-0.7$ & $-1.05$ & $-0.95$ & $-1.3$ & $-0.7$ & $-1.15$ & $-0.85$ \\

$w_a$ & \multicolumn{2}{c}{*} & \multicolumn{2}{c}{*} & $-1.73$ & $1.28$ & $-0.4$ & $0.4$ & $-0.7$ & $0.5$ & $-0.3$ & $0.3$ \\

\hline

\end{tabular}
\label{tab:emulator_ranges_extended}
\end{center}
\end{table*}

\begin{figure*}
\begin{center}
\hspace*{-0.75cm}\includegraphics[height=18.5cm,angle=270]{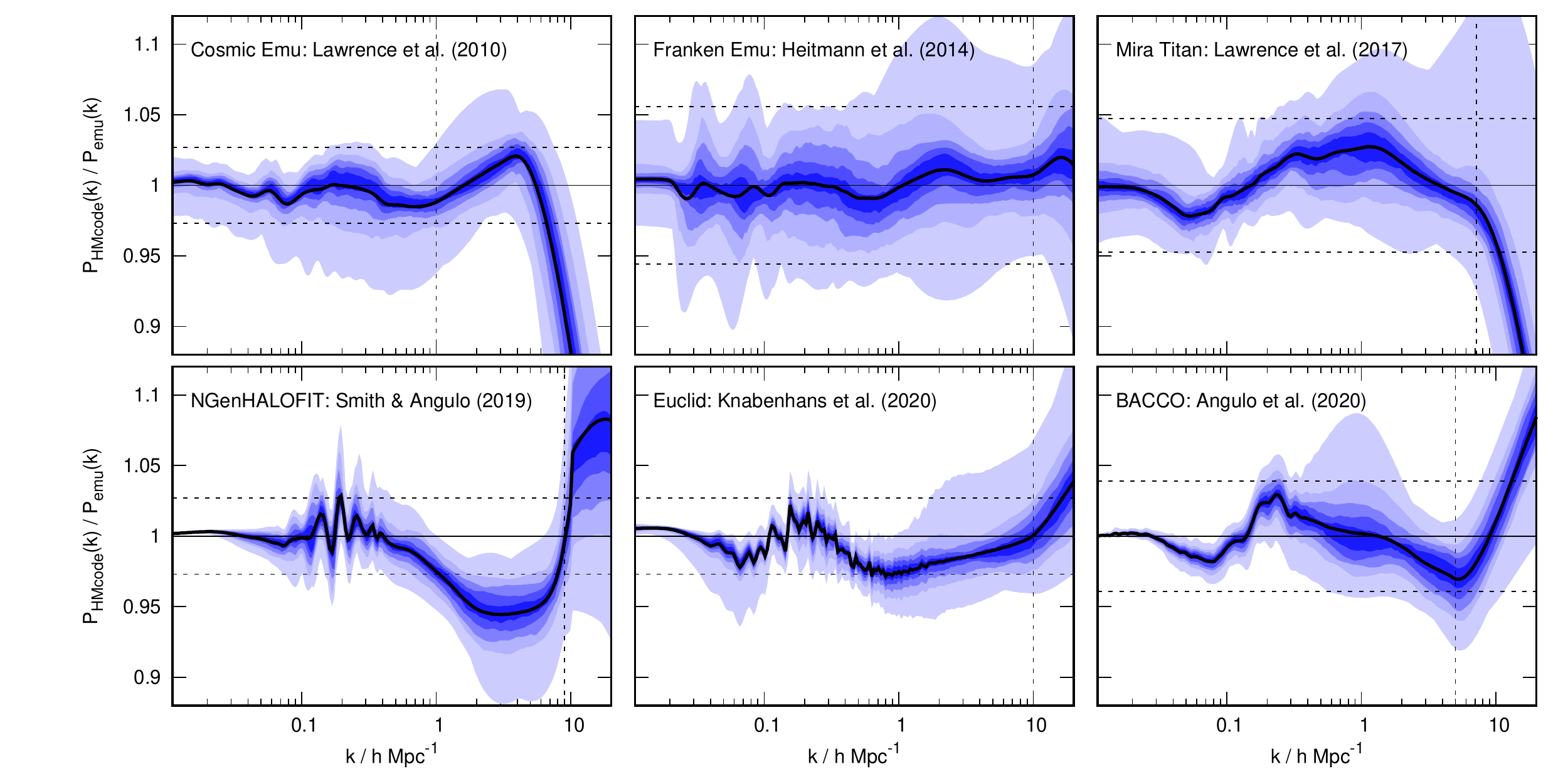}
\end{center}
\caption{Performance of \hmcodett compared to emulator predictions for $100$ (different) random cosmological models drawn from the parameter space of each emulator at $z=0$. The black line shows the mean prediction across cosmologies, while the coloured bands show percentiles, with the inner-most band containing the central $20$ per cent and then stepping out in $20$ percent steps to reach $100$ percent for the outer-most band. Vertical dashed lines show the maximum $k$ emulated, for power beyond this wavenumber we extend the emulator prediction as a power law (we would not expect this to be accurate). Dashed horizontal lines show the expected error if we assume that the 2.5 per cent \hmcodett error and that of the emulator can be added in quadrature.}
\label{fig:random_emulator_comparison}
\end{figure*}

In Fig.~\ref{fig:random_emulator_comparison} we show a comparison between the non-linear power spectrum predicted by \hmcodett and that from emulators at $z=0$: \cosmicemu \citep[1 per cent accurate;][]{Lawrence2010}, \frankenemu \citep[5 per cent accurate;][]{Heitmann2014}, \miratitan \citep[4 per cent accurate;][]{Lawrence2017}, \halofit \citep[1 per cent accurate;][]{Smith2019}, \euclid \citep[version 2; 1 per cent accurate;][]{Knabenhans2020} and \bacco \citep[3 per cent accurate;][]{Angulo2020}. In the main body of this paper we compared to power from the node cosmologies of \frankenemu and \miratitan, which we took to be essentially perfect, but here we compare to $100$ (uniform) random cosmologies drawn from the quoted cosmology range of each emulator (see Table~\ref{tab:emulator_ranges_extended}). Note that this means that $100$ cosmologies are different in each panel, because each emulator covers a different range of cosmological-parameter space. We therefore need to consider the accuracy of each emulator together with our quoted $2.5$ per cent for \hmcodett. We include the new version of \halofit in this `emulator' comparison because, within its parameter space, it calculates the non-linear power by interpolating and/or extrapolating \nbody data in a similar way to the other emulators (although they all use different techniques). However, for the \halofit comparison we set the running of the spectral index to zero; if we do not we see a large scatter between the predictions, presumably because none of the halo-model ingredients have been calibrated for cosmologies with running spectral indices. This indicates that improving the halo model for cosmologies with running spectral indices may be a fruitful direction for further research.

From the Fig.~\ref{fig:random_emulator_comparison} we see good agreement between \hmcodett and the emulators in all cases, and the agreement is within the accuracy range of each emulator if we assume that the errors add with that of \hmcodett in quadrature. The exception is with the \halofit emulator, where we see that \hmcodett systematically overpredicts the power by a few per cent for $k\sim5\iMpc$; this overprediction is not present in the other comparisons. From the comparisons with \miratitan, \euclid, and \bacco we note a systematic feature of \hmcodett to slightly under-predict the power for $k\sim0.08\iMpc$ and then slightly overpredict for $k\sim0.2\iMpc$. We suspect that this may be due to our relatively schematic treatment of these `perturbative' scales, and we note that this systematic bias vanishes at higher $z$. The larger deviations seen in the \miratitan and \bacco panels for $k\sim1\iMpc$ are all cosmologies with high neutrino fractions. The larger deviations seen in the \miratitan panel for $k\sim10\iMpc$ are all cosmologies with dark energy equations of state that lead to significant early dark energy. We note that compared to the \euclid emulator our power is systematically slightly low for $k\sim1\iMpc$, but it is systematically slightly high around the same scales when compared to \miratitan, which indicates that there may be differences between the power produced by the different simulation techniques at this scale.

Note that Fig.~\ref{fig:random_emulator_comparison} should not necessarily be used to compare the performance of the emulators themselves, as each covers a different sub-space of cosmological parameters (see Table~\ref{tab:emulator_ranges_extended}). Some parameter combinations make predictions more difficult, and \hmcode may be systematically biased in some regions of parameter space. For example, the parameter space of \miratitan covers high-mass neutrinos and dark energy models with high-$z$ equations of state that are close to zero. The near-perfect agreement at large scales with \halofit arises because this model defaults to a perturbative treatment of power at these scales, and this uses exactly the same linear spectrum as the \hmcode calculation. The agreement at large scales with \euclid and \bacco is a result of these emulators emulating a `correction function' to the linear power spectrum, whereas the other techniques emulate the non-linear power directly. For \bacco, this correction factor asymptotes to unity at large scales, so the agreement is perfect with \hmcodett because each model is then simply linear theory. The $\sim 0.5$ per cent high bias compared to \euclid arises from the correction factor not asymptoting perfectly to unity, even at large scales. In \cite{Knabenhans2020} this difference (which also exists between the different versions of the \euclid emulator) is attributed to simulation volume. 

\end{document}